\documentclass[a4paper,
               amsmath,
               amssymb,
               aps,
               floatfix,
               longbibliography,
               prd,
               reprint,
               showkeys,
               superscriptaddress]{revtex4-2}

\usepackage{cmap}

\usepackage[utf8]{inputenc}

\usepackage[T1]{fontenc}

\usepackage[american]{babel}

\usepackage[usenames, dvipsnames]{xcolor}

\usepackage{etoolbox}
\apptocmd{\sloppy}{\hbadness 10000\relax}{}{}

\AtBeginEnvironment{thebibliography}{\interlinepenalty=10000}

\newlength{\bibitemsep}\newlength{\bibparskip}
\let\oldthebibliography\thebibliography
\renewcommand\thebibliography[1]{%
    \oldthebibliography{#1}%
    \setlength{\parskip}{\bibitemsep}%
    \setlength{\itemsep}{\bibparskip}%
}
\usepackage{hyperxmp}

\usepackage[pdftex,
            unicode=true,
            colorlinks,
            urlcolor=Blue,
            linkcolor=gray,
            citecolor=gray,
            filecolor=gray,
            plainpages=false,
            pdfpagelabels]{hyperref}

\hypersetup{pdftitle={Convolutional neural networks: a magic bullet for gravitational wave detection?},
            pdfauthor={Timothy D. Gebhard, %
                       Niki Kilbertus, %
                       Ian Harry, %
                       Bernhard Schölkopf},
            pdfkeywords={Convolutional Neural Networks, %
                         Gravitational Waves, %
                         Compact Binary Coalescences, %
                         Binary Black Holes, %
                         Deep Learning, %
                         Machine Learning, %
                         Binary Classification, %
                         Tagging} }

\usepackage[noabbrev]{cleveref}

\usepackage{url}

\usepackage[all]{hypcap}

\usepackage[includefoot,
            includehead,
            bindingoffset=0cm,
            left=2cm,
            right=2cm,
            top=1.5cm,
            bottom=1.5cm]{geometry}

\usepackage{graphicx}

\usepackage[caption=false]{subfig}
\usepackage{tikz}
\usetikzlibrary{calc,
                quotes,
                positioning,
                shapes.geometric,
                decorations.markings}

\usepackage{pgfplots}
\pgfplotsset{compat=1.14}

\definecolor{color0}{HTML}{1F77B4}
\definecolor{color1}{HTML}{FF7F0E}
\definecolor{color2}{HTML}{2CA02C}
\definecolor{color3}{HTML}{D62728}
\definecolor{color4}{HTML}{9467BD}
\definecolor{color5}{HTML}{8C564B}
\definecolor{color6}{HTML}{E377C2}
\definecolor{color7}{HTML}{7F7F7F}
\definecolor{color8}{HTML}{BCBD22}
\definecolor{color9}{HTML}{17BECF}

\usepackage{libertine}
\usepackage[libertine]{newtxmath}
\usepackage[scaled=0.76]{beramono}

\usepackage{setspace}
\usepackage[activate={true, nocompatibility},
            final,
            babel=true,
            tracking=true,
            kerning=true,
            spacing=true,
            factor=1100,
            stretch=5,
            shrink=50]{microtype}
\SetTracking{encoding={*}, shape=sc}{10}

\usepackage{amsmath}
\usepackage{amsthm}
\usepackage{amssymb}

\usepackage{csquotes}

\usepackage{nicefrac}

\usepackage{siunitx}
\let\sun\odot
\DeclareSIUnit \solarmass {\ensuremath{M_\sun}}
\DeclareSIUnit \parsec {pc}
\DeclareSIUnit \year {yr}

\usepackage[inline]{enumitem}

\usepackage{textcomp}

\newcommand{\snr}{\ensuremath{\mathrm{SNR}}}

\newcommand{\eg}{e.g.,~}
\newcommand{\ie}{i.e.,~}

\newcommand{\pycbc}{\textsc{PyCBC}}

\newcommand{\OR}[1]{O#1}

\newcommand{\cross}{\ensuremath{\times}}

\makeatletter
\DeclareRobustCommand*{\escapeus}[1]{%
  \begingroup\@activeus\scantokens{#1\endinput}\endgroup}
\begingroup\lccode`\~=`\_\relax
   \lowercase{\endgroup\def\@activeus{\catcode`\_=\active \let~\_}}
\makeatother

\usepackage{tcolorbox}
\definecolor{hltextttgray}{rgb}{0.75, 0.75, 0.75}
\newtcbox{\hlbox}{nobeforeafter,
                  colframe=hltextttgray,
                  colback=hltextttgray!10,
                  boxrule=0.5pt,
                  arc=3pt,
                  boxsep=0pt,
                  left=2pt,
                  right=2pt,
                  top=1.5pt,
                  bottom=1.5pt,
                  tcbox raise base}
\newcommand{\hltt}[1]{\hlbox{\texttt{\escapeus{\vphantom{j}#1}}}}

\begin{document}

    \title{Convolutional neural networks: A magic bullet for gravitational-wave detection?}
   
    \author{Timothy D. Gebhard}
    \thanks{These two authors contributed equally.\\ Correspondence: \href{mailto:tgebhard@tue.mpg.de,nkilbertus@tue.mpg.de}{\texttt{\{tgebhard,nkilbertus\}@tue.mpg.de}}.}
    \affiliation{Max Planck Institute for Intelligent Systems, Max-Planck-Ring 4, 72076 Tübingen, Germany}
    \affiliation{Max Planck ETH Center for Learning Systems, Universitätstrasse 6, 8092 Zürich, Switzerland}

    \author{Niki Kilbertus}
    \thanks{These two authors contributed equally.\\ Correspondence: \href{mailto:tgebhard@tue.mpg.de,nkilbertus@tue.mpg.de}{\texttt{\{tgebhard,nkilbertus\}@tue.mpg.de}}.}
    \affiliation{Max Planck Institute for Intelligent Systems, Max-Planck-Ring 4, 72076 Tübingen, Germany}
    \affiliation{Engineering Department, University of Cambridge, Trumpington Street, Cambridge, CB2 1PZ, UK}
    
    \author{Ian Harry}
    \affiliation{Institute for Cosmology and Gravitation, University of Portsmouth, 1-8 Burnaby Road, Portsmouth, P01 3FZ, UK}
    \affiliation{Max Planck Institute for Gravitational Physics, Am Mühlenberg 1, 14476 Potsdam, Germany}

    \author{Bernhard Schölkopf}
    \affiliation{Max Planck Institute for Intelligent Systems, Max-Planck-Ring 4, 72076 Tübingen, Germany}

    \date{\today}

    \begin{abstract}
        In the last few years, machine learning techniques, in particular convolutional neural networks, have been investigated as a method to replace or complement traditional matched filtering techniques that are used to detect the gravitational-wave signature of merging black holes. 
        However, to date, these methods have not yet been successfully applied to the analysis of long stretches of data recorded by the Advanced LIGO and Virgo gravitational-wave observatories. 
        In this work, we critically examine the use of convolutional neural networks as a tool to search for merging black holes. 
        We identify the strengths and limitations of this approach, highlight some common pitfalls in translating between machine learning and gravitational-wave astronomy, and discuss the interdisciplinary challenges. 
        In particular, we explain in detail why convolutional neural networks alone cannot be used to claim a statistically significant gravitational-wave detection. 
        However, we demonstrate how they can still be used to rapidly flag the times of potential signals in the data for a more detailed follow-up. 
        Our convolutional neural network architecture as well as the proposed performance metrics are better suited for this task than a standard binary classifications scheme. 
        A detailed evaluation of our approach on Advanced LIGO data demonstrates the potential of such systems as trigger generators.
        Finally, we sound a note of caution by constructing \emph{adversarial examples}, which showcase interesting ``failure modes'' of our model, where inputs with no visible resemblance to real gravitational-wave signals are identified as such by the network with high confidence.
    \end{abstract}
    

    \maketitle

        \section{Introduction}
        \label{sec:introduction}

        Matched filtering techniques~\citep{Allen_2012, Babak_2013, Usman_2016, Messick_2017} have proven highly successful in discovering binary black hole coalescences from the recordings of the Advanced LIGO and Advanced Virgo gravitational-wave observatories~\citep{Aasi_2015, Acernese_2014, Abbott_2016-02-11, Abbott_2016-06-15, Abbott_2017-06-01, Abbott_2017-10-06, Abbott_2017-12-18}.
        Ten observations of merging black holes have now been made~\citep{Abbott_2018-12-16}.
        These observations have enabled population studies of the properties of stellar-mass black holes and allowed precision tests of general relativity to be carried out~\citep{Abbott_2018-12-16, Abbott_2019-01-03}.
        The most important observation to date was arguably the detection of a binary neutron star inspiral together with a gamma-ray burst and other electromagnetic counterparts~\citep{Abbott_2017-10-16a, Abbott_2017-10-16b}.
        This detection heralds the era of multimessenger gravitational-wave astronomy, has yielded an independent measurement of Hubble's constant, and probed the behavior of matter at the core of neutron stars~\citep{Abbott_2017-10-16, Abbott_2018-10-15}.
        
        Additional observatories in Japan and India are expected to become operational in the next five years forming an evolving detector network capable of observing hundreds of sources every year~\citep{Aso_2013, Abbott_2018-04-26}. 
        These sources will need to be rapidly observed, localized in the sky and this information quickly disseminated to electromagnetic partners to maximize the chance of multimessenger observations~\citep{Abbott_2018-04-26}.
        This requires reliable, real-time identification of potential compact binary coalescences (CBCs) to provide a time window and basic parameter estimate for slower, but more accurate Bayesian inference techniques to follow-up~\citep{Veitch_2015, Singer_2016}.
        However, current matched filtering techniques are computationally expensive, with the computational cost scaling as a function of the broadness of the detector's sensitivity curve and the number of observatories; both of which are expected to increase in the coming years~\citep{Abbott_2018-04-26}.
        
        In this work, we investigate whether some of these challenges can efficiently be overcome by using deep convolutional neural networks (CNNs).
        CNNs are a machine learning technique that has been employed successfully on a wide variety of tasks, including image classification~\citep{LeCun_1989, LeCun_1998, Krizhevsky_2012}, natural language processing~\citep{Kim_2014} and audio generation~\citep{VanDenOord_2016}.
        In the physics community, an early application of CNNs was \cite{Zhu_2014}; \citet{Carleo_2019} provide a review of recent developments in this direction.
        In particular, CNNs have also been studied in the literature as a tool for gravitational-wave searches, and previous works have shown that they can indeed be effectively applied to this problem when treating it as a binary (i.e., two-class) classification task~\citep{George_2016, Gabbard_2018}.
        
        However, despite these promising preliminary results, we believe that the precise role that machine learning can play within the larger scope of CBC searches and practical multimessenger gravitational-wave astronomy has not yet been assayed in sufficient detail.
        The main goal of this work is, therefore, to carefully and realistically analyze the practical potential of using CNNs to search for GWs from CBCs. 
        Here, we pay particular attention to realistic data generation, an appropriate, task-specific architecture design and adequately chosen performance metrics.
        This results in the following main contributions:
        
        \begin{enumerate}
            \item We provide an in-depth analysis of the challenges one may expect machine learning to solve within the scope of a search for GWs from CBCs, and also discuss their limitations in replacing matched filtering or Bayesian parameter estimation techniques.
            \item We extend the existing, binary classification-based approach of using CNNs to also handle inputs of varying length.
            This requires the introduction of new task-specific performance metrics, which we discuss and relate to the existing metrics.
            \item We highlight potential challenges and subtle pitfalls in the data generation process that may lead to unfair comparisons.
            To facilitate further research and reproducibility in this area, we release the data generation workflow we have developed as a reusable open source software package.
            \item Finally, the empirical results of our architecture indicate that deep convolutional neural networks are a powerful supplement to the existing pipeline for fast and reliable trigger generation.
            However, we also demonstrate that---like most deep neural networks---our architecture is also prone to \emph{adversarial attacks}: We can construct inputs with no visible resemblance to gravitational-wave signals that are nevertheless identified as such by the model.
        \end{enumerate}
        
        As a key aspect of this work, we aim to foster communication and understanding between disciplines:
        On the one hand, we hope to help physicists less acquainted with deep learning techniques understand the strengths and limitations of such methods in gravitational-wave searches and gain intuition towards how they function in this context.
        Simultaneously, for machine learning experts, we explicitly highlight some problem-specific subtleties---ranging from data generation to model architecture design and meaningful evaluation metrics---to help them to circumvent tempting pitfalls.
        
        The rest of this paper is structured as follows.
        In \cref{sec:problem-setup-and-related-work}, we revisit matched filtering (with a focus on the implementation by \pycbc{}).
        Furthermore, we discuss the existing literature on using CNNs in the context of gravitational-wave searches.
        In \cref{sec:going-beyond-binary-classifcation}, we then continue by reviewing the previously used binary classification framework more principally, and discuss for which specific tasks CNNs may be useful and for which their output is insufficient.
        Consequently, after introducing our carefully designed data generation procedure and the corresponding open source software package in \cref{sec:data-generation-process}, we suggest a fully convolutional network architecture suited for gravitational-wave trigger generation in streaming data in \cref{sec:model-and-training-procedure}.
        This architecture naturally gives way to novel performance metrics, which we develop in \cref{sec:performance-metrics}, where we also explain their benefits and relation to current standard metrics.
        In \cref{sec:experiments-and-results}, we present and discuss the results of our model together with a note of caution concerning adversarial examples, highlighting the still not well-understood and unsettling brittleness of deep neural networks.
        Finally, we conclude with a summary and outlook in \cref{sec:discussion-and-conclusion}.

        \section{Problem setup and related work}
        \label{sec:problem-setup-and-related-work}
        
        Observing compact binary coalescences has always been one of the primary goals of gravitational-wave astronomy.
        To date, searches for such systems rely on matched filtering using a large template bank (\ie a set of simulated waveforms covering a carefully chosen parameter space).
        In the first part of this section, we will describe matched filtering with a specific focus on the implementation provided by the \pycbc{} software package~\citep{Usman_2016, PyCBC}.
        We explain the necessary components for a statistically sound search procedure and explain what it means to \enquote{detect} a gravitational wave.
        Readers familiar with the matched filtering search pipeline may wish to skip parts~\ref{subsec:mf}, \ref{subsec:pycbc}, and \ref{subsec:injections}.
        In part~\ref{subsec:existing}, we then review the existing work using convolutional neural networks for gravitational-wave searches.

        \subsection{Matched filtering-based searches}
        \label{subsec:mf}

        \citet{Schutz_1999} vividly describes the intuition behind the matched filtering technique as follows:
        \enquote{Matched filtering works by multiplying the output of the detector by a function of time (called the template) that represents an expected waveform, and summing (integrating) the result. If there is a signal matching the waveform buried in the noise then the output of the filter will be higher than expected for pure noise.}
        
        In the following, we will formalize this idea mathematically in order to provide the necessary background for a comparison between matched filtering and the outputs of deep learning-based systems later on.
        Readers interested in further details are referred to the excellent overview of matched filtering in the context of the LIGO and Virgo collaborations by \citet{Caudill_2018} (and references therein).

        The fundamental assumption of matched filtering is that the \emph{strain}~$s(t)$ measured by the interferometric detector is made up of two \emph{additive} components, namely the \emph{instrument noise}~$n(t)$ and the (astrophysical) \emph{signal}~$h(t)$:
        \begin{equation}
            s(t) = n(t) + h(t)
        \end{equation}
        For a given power spectral density $S_n$ of $n$, we can then quantify the agreement between a given template $T(t)$ in the template bank and the recorded strain $s(t)$ at a time $t_0$ by computing the \emph{signal-to-noise ratio} (SNR).
        For an appropriate choice of normalization, the matched filtering signal-to-noise ratio is given by:
        \begin{equation}
            \snr(t_0) := \int_{-\infty}^{\infty} \frac{\tilde{s}(f) \cdot \tilde{T}^*(f) \cdot e^{2 \pi i f t_0}}{S_n(f)} \, \text{d}f \:,
        \end{equation}
        where the tilde denotes the Fourier transform.
        For stationary Gaussian noise it can be shown that---by design---the SNR is indeed the optimal detection statistic for finding a signal $h(t)$ if the time-reversed template $T(-t)$ is equal to the signal~\cite{Allen_2012}.
        This is called the \emph{matched} filter.
        In practice, the template bank should therefore contain accurate simulated waveforms that cover the space of expected signals in the recorded data sufficiently densely.
        Computing the SNR for every waveform in the template bank and applying a threshold then produces a list of candidate event times.

        In reality, however, the data is usually neither stationary nor exactly Gaussian.
        One particular challenge to the data analysis are so called \emph{glitches}.
        Glitches are nonstationary noise transients, which comprise a range of different short-time phenomena that affect the quality of the data measured by the detectors. 
        They occur frequently, at rates up to several times per minute~\cite{Abbott_2016-06-06}. 
        Some of these effects are well understood, such as the signature of scattered light in the beam tube; others, however, remain enigmatic.
        For example, a certain common type of glitch named \enquote{blip}, whose origin is only poorly understood, tends to mimic the signals that one would expect from the merger of two intermediate-mass black holes, thus limiting the sensitivity for this kind of event~\cite{Cabero_2019}.

        As a consequence of these non-Gaussian and nonstationary effects, the real distribution of the SNR (and thus the threshold value) is not known and must be determined empirically in order to obtain calibrated statistical results from the computed SNR.
        \citet{Allen_2012} provide a detailed account of the merits and challenges of matched filtering in practical gravitational-wave searches.

        \subsection{The \pycbc{} search pipeline}
        \label{subsec:pycbc}

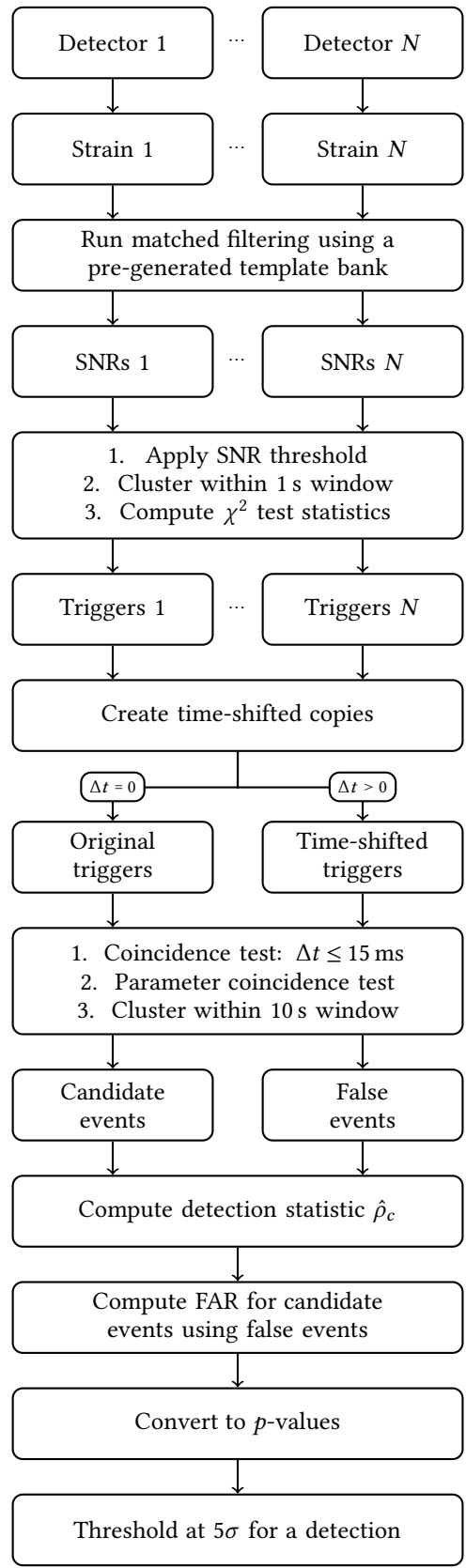
\begin{figure}
            \centering
            \begin{tikzpicture}

                \tikzstyle{halfblock} = [rectangle, draw, fill=white, minimum width=2.8cm, text width=2.5cm, inner sep=1mm, minimum height=1cm, text centered, rounded corners, anchor=center, thick, align=center]
                \tikzstyle{wideblock} = [rectangle, draw, minimum width=6.3cm, fill=white, text width=6.1cm, inner sep=1mm, minimum height=1cm, text centered, rounded corners, anchor=center, thick, align=center]
                \tikzstyle{arr} = [draw, thick, ->]

                \node [halfblock] (detector_1) at (-1.75, -0.5) 
                    {Detector 1};
                \node [] at (0, -0.5) 
                    {$\scriptstyle\cdots$};
                \node [halfblock] (detector_n) at (+1.75, -0.5) 
                    {Detector $N$};
                \node [halfblock] (strain_1) at (-1.75, -2.0) 
                    {Strain 1};
                \node [] at (0, -2.0) 
                    {$\scriptstyle\cdots$};
                \node [halfblock] (strain_n) at (+1.75, -2.0) 
                    {Strain $N$};
                \node [wideblock] (matched_filtering) at (0.0, -3.5) 
                    {Run matched filtering using a pre-generated template bank};
                \node [halfblock] (snr_1) at (-1.75, -5.0) 
                    {SNRs 1};
                \node [] at (0, -5.0) 
                    {$\scriptstyle\cdots$};
                \node [halfblock] (snr_n) at (+1.75, -5.0) 
                    {SNRs $N$};
                \node [wideblock, minimum height=1.5cm, align=center] (snr_threshold) at (0.0, -6.75)
                    {1. Apply SNR threshold\\
                     2. Cluster within \SI{1}{\second} window\\
                     3. Compute $\chi^2$ test statistics};
                \node [halfblock] (triggers_1) at (-1.75, -8.5) 
                    {Triggers 1};
                \node [] at (0, -8.5) 
                    {$\scriptstyle\cdots$};
                \node [halfblock] (triggers_n) at (+1.75, -8.5) 
                    {Triggers $N$};
                \node [wideblock] (time_shifts) at (0.0, -10.0) 
                    {Create time-shifted copies};
                \node [halfblock, minimum width=2.8cm, text width=2.5cm] (original_triggers) at (-1.75, -12.00) 
                    {Original\\ triggers};
                \node [halfblock, minimum width=2.8cm, text width=2.5cm] (shifted_triggers) at (+1.75, -12.0) 
                    {Time-shifted\\ triggers};
                \node [wideblock, minimum height=1.5cm, align=center] (coincidence_test) at (0.0, -13.75)
                    {1. Coincidence test: $\Delta t\,{\leq}\,\SI{15}{\milli\second}$\\
                     2. Parameter coincidence test\\
                     3. Cluster within \SI{10}{\second} window};
                \node [halfblock, minimum width=2.8cm, text width=2.5cm] (candidate_events) at (-1.75, -15.5) 
                    {Candidate\\ events};
                \node [halfblock, minimum width=2.8cm, text width=2.5cm] (false_triggers) at (+1.75, -15.5) 
                    {False\\ events};
                \node [wideblock] (det_stat) at (0.0, -17.0)
                    {Compute detection statistic $\hat{\rho}_c$};
                \node [wideblock] (far) at (0.0, -18.5)
                    {Compute FAR for candidate\\ events using false events};
                \node [wideblock] (pvalue) at (0.0, -20.0)
                    {Convert to $p$-values};
                \node [wideblock] (detection) at (0.0, -21.5)
                    {Threshold at $5\sigma$ for a detection};

                \draw [arr] (detector_1.south) -- (strain_1.north);
                \draw [arr] (detector_n.south) -- (strain_n.north);
                \draw [arr] (strain_1) -- ([xshift=-1.75cm]matched_filtering.north);
                \draw [arr] (strain_n) -- ([xshift=+1.75cm]matched_filtering.north);
                \draw [arr] ([xshift=-1.75cm]matched_filtering.south) -- (snr_1.north);
                \draw [arr] ([xshift=+1.75cm]matched_filtering.south) -- (snr_n.north);
                \draw [arr] (snr_1.south) -- ([xshift=-1.75cm]snr_threshold.north);
                \draw [arr] (snr_n.south) -- ([xshift=+1.75cm]snr_threshold.north);
                \draw [arr] ([xshift=-1.75cm]snr_threshold.south) -- (triggers_1.north);
                \draw [arr] ([xshift=+1.75cm]snr_threshold.south) -- (triggers_n.north);
                \draw [arr] (triggers_1.south) -- ([xshift=-1.75cm]time_shifts.north);
                \draw [arr] (triggers_n.south) -- ([xshift=+1.75cm]time_shifts.north);
                \draw [arr] (time_shifts.south) |- ++(0, -0.5) -| (original_triggers.north) node [anchor=mid, draw, fill=white, rounded corners, midway, font=\scriptsize] 
                    {$\Delta t$~\raisebox{0.3mm}{$\scriptscriptstyle =$}~$0$};
                \draw [arr] (time_shifts.south) |- ++(0, -0.5) -|  (shifted_triggers.north) node [anchor=mid, draw, fill=white, rounded corners, midway, font=\scriptsize] 
                    {$\Delta t$~\raisebox{0.3mm}{$\scriptscriptstyle >$}~$0$};
                \draw [arr] (original_triggers.south) -- ([xshift=-1.75cm]coincidence_test.north);
                \draw [arr] (shifted_triggers.south) -- ([xshift=+1.75cm]coincidence_test.north);
                \draw [arr] ([xshift=-1.75cm]coincidence_test.south) -- (candidate_events.north);
                \draw [arr] ([xshift=+1.75cm]coincidence_test.south) -- (false_triggers.north);
                \draw [arr] (candidate_events.south) --  ([xshift=-1.75cm]det_stat.north);
                \draw [arr] (false_triggers.south) -- ([xshift=+1.75cm]det_stat.north);
                \draw [arr] (det_stat.south) -- (far.north);
                \draw [arr] (far.south) -- (pvalue.north);
                \draw [arr] (pvalue.south) -- (detection.north);

            \end{tikzpicture}
            \caption{Flowchart of the \pycbc{} search pipeline, which shows the full process of going from the recordings of the different observatories to the detection of a gravitational wave.}
            \label{fig:pycbc-flowchart}
        \end{figure}

        To understand the crucial components of a \emph{full} search (which ideally results in a \emph{detection}), we now outline the current \pycbc{} search pipeline~\cite{Usman_2016}.
        The different steps of the search procedure are also illustrated schematically as a flowchart in \cref{fig:pycbc-flowchart}.

        In a first step, a template bank containing simulated waveforms that cover the parameter space of interest is constructed; typically using the simulation routines provided by \textsc{LALSuite} \cite{LALSuite}, the central codebase that implements all waveform models used in Advanced LIGO and Advanced Virgo analyses.
        For more technical details we refer the reader to, for example, \citet{Capano_2016}.

        This template bank is then used to compute an $\snr$ time series for every possible combination of templates and recordings (\ie we match every template with every observatory).
        We then find the times of peaks within all these $\snr$ time series that exceed a certain pre-defined threshold.
        Next we cluster these times to keep only the times of largest SNR within a 1-second window and then store the remaining times alongside the parameters of the template that caused the match.
        Each of these recordings is called a \emph{trigger}.

        Consequently, we obtain a list of single detector triggers for each observatory independently.
        Furthermore, a set of signal consistency tests---$\chi^2$ tests---are computed for every trigger, which help to discriminate between real events and triggers that were caused by noise transients~\citep{Allen_2005}.
        More precisely, these $\chi^2$-test values are used to compute a re-weighted single detector $\snr$ which serves as a ranking statistic.
        In a subsequent stage, several coincidence tests (for both the event time and the estimated event parameters) are conducted: 
        the single detector triggers are combined if the same template matched at compatible times (\ie within light distance of each other) in all detectors.
        The resulting coincident triggers are called \emph{candidate events}.
        Finally, each candidate event is assigned a combined ranking statistic, informally called \emph{loudness}, which is computed from the parameters of the triggers in each observatory.
        The precise mathematical definitions of the individual and combined ranking statistics are hand-tuned and regularly adjusted (see, for example, \citet{Nitz_2017-11-07} or \citet{Nitz_2018-01-12}).

        Note that while the loudness is designed to intuitively correspond to our confidence of the candidate being a real event (higher scores indicating higher confidence), the raw numerical values have no significance. 
        Instead, we are interested in the relative ordering of the candidate events that is induced by the loudness score.
        To claim a \emph{detection}---that is, to say that a candidate event with a given loudness in fact corresponds to a true gravitational-wave signal---we must perform the following statistical test: within our model assumptions, what is the probability that we observe this loudness purely by chance, if in reality there is no gravitational-wave signal present?
        This probability measures the statistical significance of the detection, that is, the confidence with which we can reject the null hypothesis, namely \enquote{there was no real signal in the data}.

        At this point, it is crucial to contrast this with deep learning based machine learning classifiers.
        The output of such classifier on a single example---for example, from a softmax or sigmoid output layer---is also between $0$ and $1$ and thus at times interpreted as a probability.
        However, these \enquote{probabilities} only reflect the \enquote{degree of confidence} of the network regarding its prediction. 
        Therefore, they must not be interpreted as the statistical significance of a detection (see also \cref{sec:going-beyond-binary-classifcation}).

        In \pycbc{}, the probability of obtaining a given loudness from only noise is estimated via frequentist inference over a given time period.
        To this end, a matched filtering search is performed on a recording of given length $T$ that is known to not contain any gravitational-wave signals.
        We then count the number of resulting candidate events that are at least as loud as the candidate event.

        To obtain data that is guaranteed to not contain any gravitational-wave signals but still shares characteristics of real detector recordings, \pycbc{} makes use of time shifts.
        It shifts the recordings of the detectors relative to each other by a time period that is larger than the light travel time between them (see again \cref{fig:pycbc-flowchart} for where this fits in the pipeline).
        Assuming that gravitational waves above the detection threshold of the instrument are sparse in time (\ie further apart than the time shift), this ensures that no real signal will pass the coincidence tests and give rise to a candidate event.
        Instead, any candidate event found for a time-shifted input must be due to triggers caused by the random detector noise.
        Therefore, the loudness scores of candidate events found in time-shifted data can be used to estimate the frequency of false positives. 
        This further allows us to derive false alarm rates for candidate events in the non time-shifted data and ultimately assign a statistical significance to a claimed detection.
        For a slightly more detailed yet compact description of how to estimate these probabilities in practice, we again refer to \citet{Caudill_2018}.

        \subsection{Injections}
        \label{subsec:injections}

        To conclude this introduction to the existing search pipeline, we note that due to the relatively small number of events detected so far, a proper performance evaluation of \emph{any} search approach hinges on so called \emph{injections}.
        An injection is a simulated waveform that is added into a piece of background noise (either synthetic or real) to emulate a real gravitational-wave signal as it would be observed by an actual detector.
        The search performance can then be evaluated by searching for a large variety of such injections added to the recorded strain data.
        Because in this case we know the precise location of the injections, we have access to the ground truth required to evaluate the detection rate and false alarm rate of the search pipeline for a given template bank, real recordings, and injections.

        In the previous paragraphs, we have glanced over the fact that we can only compute false alarm probabilities and detection rates \emph{within our model assumptions}.
        These assumptions include---among other factors---the parameter ranges and distributions of simulated waveforms both for the template bank and injections.
        Since the true physical distribution of gravitational-wave sources in the Universe (not only in terms of location, but also in terms of the parameters of their constituents) is unknown, these choices will not only affect how the obtained performance results transfer to real searches, but also influence the sensitivity towards various sources. 
        In \cref{sec:data-generation-process}, we comment on this in a little more detail.
        However, a full discussion of how to properly incorporate such ad hoc choices in the statistical analysis of the method is beyond of the scope of this work.

        \subsection{Existing CNN-based approaches}
        \label{subsec:existing}

        The idea of using convolutional neural networks (CNNs) to process time series information goes back to the early days of deep learning itself, more than twenty years ago \citep{LeCun_1995}.
        Ever since, the community has established CNNs as one of the major work horses for processing images as well as time series data like audio (or various time-frequency representation thereof), which is structurally similar to the strain data produced by gravitational-wave observatories.
        CNNs have been particularly successful in supervised classification or regression tasks, where they are typically trained to map inputs in $\mathbb{R}^d$---for example, images of a fixed resolution or fixed-length audio snippets---to either a finite set of labels (classification) or a typically low-dimensional real vector (regression).

        All previous work applying convolutional neural networks to the detection of gravitational-wave signals in interferometric detector data has adopted a classification-based approach.
        \citet{George_2016} generate white Gaussian noise examples with a fixed length of \SI{1}{\second} and, for a subset of them, add simulated gravitational-wave signals from binary black hole mergers similar to the injections in the \pycbc{} search.
        The maximum of the signal (which corresponds to the coalescence time) is randomly located in the last quarter of the sample.
        Using these data, they train a deep neural net, consisting of a common combination of convolutional and fully-connected layers with a final sigmoid layer, to output a value between $0$ and $1$, indicating the confidence of the network about the absence or presence of a gravitational-wave signal in each \SI{1}{\second} example.
        The network output can be thresholded to obtain a binary response.
        In addition, they train a second neural network, which estimates some basic parameters of the corresponding binary merger whenever the first network claims to have found a signal.
        In this setup, the CNN's task is to detect non-Gaussianities of a specific form in white Gaussian noise, where the non-Gaussianities fall within a specific region of the input snippet.

        In later works, they also evaluate this method on \SI{1}{\second} snippets of real LIGO recordings, and on an enlarged dataset which also includes waveforms for binary black hole mergers with precessing spins and nonvanishing orbital eccentricities \citep{George_2018a,George_2018b}.
        Longer samples are processed by a sliding-window approach: 
        recordings are split into overlapping \SI{1}{\second}-windows to each of which the trained network is applied.
        Multiple detectors are accounted for by processing each recording separately first and then combining the binary outputs at each time via a logical \textsc{and} function.
        Notably, the authors suggest that their method can be used for gravitational-wave detection as well as parameter estimation and that it beats matched filtering in terms of errors and computational efficiency while retaining similar sensitivity \citep{George_2018b}.
        We will explain in \cref{sec:going-beyond-binary-classifcation} why we believe that a more careful and nuanced interpretation of such claims is essential to understanding the practical merits of CNN based approaches.

        \citet{Gabbard_2018} employ a similar approach:
        the authors also use a deep neural network consisting of both convolutional and fully connected layers to perform a binary classification task on \SI{1}{\second} samples of Gaussian noise which either do or do not contain a simulated GW signal.
        The focus of their work, however, is the comparison with matched filtering.
        They conclude that their method is indeed able to closely reproduce the results of a matched filtering-based search on these \SI{1}{s} samples.

        A somewhat different approach was presented by \citet{Li_2017}.
        In their method, they use a wavelet packet decomposition to preprocess the data before feeding it into a convolutional neural network, which then operates on a frequency representation.
        They also work with a sliding-window approach to apply their network to samples of variable length.
        Ultimately, the practical conclusions of their work are limited by the fact that they use Gaussian noise for the background and an unrealistically simplified damped sinusoid as an analytical waveform model.

        Finally, there is also a growing body of work which uses CNNs for various tasks that are different from but related to a gravitational-wave \emph{search}, such as \emph{glitch classification} (\eg~\cite{Zevin_2017, Bahaadini_2017, Razzano_2018, Bahaadini_2018, Coughlin_2019}) or \emph{parameter estimation} (\eg~\cite{Shen_2019}).
        Furthermore, \citet{Dreissigacker_2019} recently presented a proof-of-principle study on using convolutional neural networks to search for \emph{continuous} gravitational waves.

        \section{Going beyond binary classification}
        \label{sec:going-beyond-binary-classifcation}

        In this section, we develop our main conceptual contributions, namely that 
        (a) convolutional neural networks are not suited to claim statistically significant detections of gravitational waves, however,
        (b) they can still be useful tools for real-time trigger generation.

        Our core argument for claim (a) hinges on the fact that the  \enquote{false alarm rate} which can be derived from machine learning-based classifiers is directly linked to the training dataset.
        As a consequence, there is only a single significance level that one can assign to every claimed detection, without being able to distinguish particularly loud events from fainter ones.
        Additional difficulties stem from the fact that in a real search, the task at hand is not to perform binary classification on fixed-length examples, but to identify the temporal location of potential signals in time series data of arbitrary length, or even in streaming data.
        The significance level obtained in the example-based binary classification setup does not transfer easily to sliding-window based approaches for streaming data.

        To substantiate (b), we highlight the benefits of CNNs in terms of computational complexity and devote the remaining sections of this paper to developing a modified CNN architecture which can overcome many of the pitfalls of the binary classification approach.

        \subsection{True / false positive rate and class imbalance}
        \label{subsec:true-false-positive-rate-and-class-imbalance}

        Standard performance metrics for classification tasks are the \emph{true positive rate} (TPR; also called \emph{recall}) and the \emph{false positive rate} (FPR), which are defined as:

        \begin{align*}
            \text{True Positive Rate (TPR)} &:= \frac{\text{TP}}{\text{TP} + \text{FN}} \,, \\
            \text{False Positive Rate (FPR)} &:= \frac{\text{FP}}{\text{FP} + \text{TN}}\,.
        \end{align*}

        Here, TP are true positives (\ie examples correctly classified as positives), FP are false positives (\ie examples falsely classified as positive; Type I error), TN are true negatives (\ie examples correctly classified as negative) and FN are false negatives (\ie examples falsely classified as negative; Type II error).

        Indeed, all previous comparisons of CNNs use a binary classification framework and compare the true positive rate at fixed false positive rate directly to matched filtering results at a given false alarm rate~\citep{George_2018a, George_2018b, Gabbard_2018}.
        To obtain this measure in practice, for threshold-based binary classifiers, one usually sweeps the threshold from $0$ to $1$, recording the true positive rate and the false positive rate for each threshold value to produce the receiver operator characteristic (ROC) curve, that is, the true positive rate over the false positive rate.
        Since the false positive rate is maximal for threshold $0$ and minimal (zero) for threshold $1$, we can then simply read off the true positive rate for any given false positive rate.
        However, there is a subtle difference between the generalization properties of this population level false positive rate and the false alarm rate in matched filtering.

        Intuitively, we may interpret the CNN as an implicit abstract representation of all the examples---with and without simulated waveforms---which it has seen during training.
        In that sense it does not directly capture a compressed version of the template bank alone, but the entire training distribution including the ratio of positive and negative examples.
        Therefore, unlike matched filtering, the network's output on new inputs depends also on the relative frequencies of positive and negative examples in the training set and the above performance measures only transfer to unseen examples following the exact same distribution.
        Consequently, performance evaluations of CNNs on the training distribution (many examples with injections) do not transfer to the test distribution (real recordings with few signals) as is the case for matched filtering, where the output depends only on the template bank.
        For efficient and stable training, the number of positive and negative examples should be on a similar order of magnitude, which is a clear misrepresentation of the true distribution and calls for caution when interpreting the FPR on hand-crafted training or validation sets as false alarm probability in a full search on real data.
        We note that in \cite{George_2018b}, the authors have computed an estimate of their FPR by applying their trained network to a continuous stretch of real LIGO data.

        \subsection{Performance vs. detection}
        \label{subsec:performance-vs-detection}

        The core task of gravitational-wave searches is not a population-level performance rating of the search pipeline on synthetic data, but to ascertain the \emph{individual} statistical significance of a given candidate event.
        Hence, we must ask ourselves the question: 
        What would be our level of confidence that there is a real event in the data when a binary classifier outputs a $1$?
        Here is the problem: If we were to use the false positive rate as a level of confidence for a claimed detection of the CNN (output $1$), we would assign the same confidence to every candidate! 
        In particular, we would have no way of distinguishing particularly significant detections from fainter ones.
        This is due to the fact that the false positive rate is a statistic of the network output on the entire dataset, not any given example.
        Furthermore, as described above, the interpretation of the false positive rate as a confidence is only valid if the test distribution (actual detector recordings) comes in well-defined, distinct fixed-length examples which follow the same distribution (including the frequencies of positive and negative examples) as the training set.
        Therefore, while the false positive rate may seem like a tempting, convenient measure for the false alarm probability of CNNs, it must not be interpreted as a statistical significance.
        Consequently, CNNs alone cannot be used to properly claim gravitational-wave detections.

        \subsection{Classification vs. tagging}
        \label{subsec:classification-vs-tagging}

        In a real search, we must identify and annotate those parts of an arbitrarily long input time series that contain a signal.
        The existing works extend the binary classification-based approach to longer inputs via a sliding window approach.
        In addition to the fixed input size of the classifier, this requires yet another parameter choice, namely the step size of the sliding window.

        Both of these parameters influence the performance metrics directly and in ways that are hard to interpret.
        First, the tempting conversion of \enquote{FPR $\times$ example length $=$ temporal rate of false positives} becomes invalid due to the overlap between neighboring windows.
        Second, depending on the step size of the sliding window, waveforms may lie only partially within the input window, which can then not be labeled as one or zero in a principled fashion.
        Moreover, there is no natural interpretation of the sequence of outputs.
        For example, assume the CNN outputs the sequence $1 - 1 - 0 - 1 - 1 - 0 - 1$, where the coalescence happens roughly at the center value.
        How should these labels be counted as true (false) positives (negatives)?
        The interpretation would perhaps also depend on the time step, that is the \emph{temporal resolution}, and the window size.
        Finally, while a high temporal resolution (small step size) would be desirable in order to localize the signal in time, it also leads to computational redundancy as we will further elaborate in \cref{sec:model-and-training-procedure}.
        
        All in all, the metrics derived from the example-based binary classification setup do not easily transfer to the sliding window approach on streaming data; a fact which has largely been overlooked in the literature so far.

        \subsection{Overfitting}
        \label{subsec:overfitting}

        We have seen that in the example-based approach, we cannot easily process inputs with partially contained waveforms.
        Previous works have therefore positioned injections only in specific regions within the examples, usually such that the coalescence is located towards the end.

        Deep learning systems are known to pick up unintentional quirks in the training data which correlate with the labels.
        This can result in an undesirable behavior called \emph{overfitting}, where a classifier learns to perform well on training data, but fails on new examples in the real application.
        In the above example, the CNN may overfit on the location of the coalescence within the training examples.
        In particular, the final, fully connected layer(s) can learn location-sensitive features.
        Since the coalescence is the most pronounced part of the waveform, if it is always located in the same region, a network containing fully connected layers may focus exclusively on high amplitude, high frequency oscillations in this region, ignoring other parts of the input.

        One crucial measure to avoid overfitting is to make the training set as representative as possible of the context in which the model will be deployed to reduce its potential to adapt to irrelevant characteristics of the training data.
        In \cref{sec:data-generation-process} and~\cref{sec:model-and-training-procedure}, we discuss a data generation process and network architecture that pay particular attention to minimizing the danger of overfitting.

        \subsection{Use-case for deep learning}
        \label{subsec:use-case-for-deep-learning}

        To conclude this section, let us discuss how CNNs can still complement matched filtering-based searches (instead of replacing them).
        Looking into the future, various upcoming challenges of matched filtering concern growing computational needs.
        For example, as more detectors come online, the computational complexity of matched filtering scales at least linearly in the number of detectors (recall that the search for triggers is performed independently for each detector first).
        Moreover, this trigger generation scales linearly also in the number of waveforms in the template bank.
        As template banks grow, matched filtering becomes increasingly expensive, causing real-time online trigger generation to become computationally challenging and prohibitive.

        Such computational considerations are a key part of the motivation to look into alternative search methods in the first place.
        Convolutional neural networks are natural candidates, because inference---evaluating the network on new strain data after it has been trained---can be substantially faster than matched filtering.
        Our architecture (see \cref{subsec:model-architecture}) scales to an arbitrary number of detectors with almost no computational overhead.
        Furthermore, once an architecture is fixed, it can in principle be trained on any distribution of simulated waveforms.
        Thus, we can view the network training as building an abstract, constant size representation of the template bank.
        Note that the computational cost of inference is independent of the size of the training data.
        The expensive training of the network is performed only once up front.

        The benefit of fast inference of CNNs---they analyze detector recordings much faster than real-time---makes them natural candidates for trigger generators.
        Real-time alarms can provide useful hints for follow up searches of electromagnetic counterparts as well as for focused analysis with matched filtering and Bayesian parameter estimation~\citep{Nitz_2018-07-30}.
        Arguably, a straightforward extension to also provide rough first parameter estimates could further decrease the computational cost of subsequent analysis by narrowing down the parameter space.

        Moreover, while CNNs do not enjoy theoretical guarantees for stationary Gaussian data like matched filtering, one may speculate that they can, in principle, incorporate mechanisms to better deal with common non-Gaussianities in the data by learning internal models not only of waveforms, but also of transient glitches.
        Testing and quantifying this hypothesis is left for future work.

        In the remainder of this work, we develop a promising proof of concept implementation for such a use-case that avoids many  pitfalls presented earlier in this section.

        \section{Data Generation Process}
        \label{sec:data-generation-process}
        
        In this section, we describe the steps we have taken to generate realistic, synthetic data which can be used to train and evaluate a CNN-based model.
        We discuss our design choices and explain steps where we found a need to compromise between realistically modeling physics on the one hand and the requirements for efficient and reliable machine learning on the other hand.
        For reasons of transparency and reproducibility, as well as to foster further research in the area, we have made our data generation code publicly available online at \cite{Gebhard_2019a}.

        \subsection{Choice of background data}
        \label{subsec:choice-of-background-data}

        When choosing background data, one has essentially two options:
        \emph{simulated} Gaussian noise, which is then \emph{colored} using the power spectral density (PSD) of the detectors, or actual detector recordings (in which the existing matched filtering pipeline did not find any gravitational-wave signals). 
        While the first option yields background data that has on average the correct frequency distribution, it will not contain glitches.
        However, as discussed before, glitches are one of the major challenges for the data analysis.
        Therefore, we have decided to use real LIGO recordings from the first observation run (\OR1) to emulate the background noise. 
        \OR1 included the first three discoveries of gravitational waves:
        GW150914, GW151012 and GW151226~\cite{Abbott_2016-02-11, Abbott_2016-06-15, Abbott_2018-12-16}.
        The exact detector configuration during \OR1 is described in detail in~\cite{Abbott_2016-03-31, Martynov_2016, Abbott_2017-03-28}.

        The data from \OR1 is publicly available through the Gravitational Wave Open Science Center (GWOSC; see also~\cite{Vallisneri_2015, GWOSC_2018}).
        In our study, we limited ourselves to a subset of the data, specified by the following criteria:
        \begin{itemize}
            \item \textbf{Data available:} Both H1 and L1 must have data available (due to different times when the detectors are operating, this is not always the case).
            \item \textbf{Minimum data quality:} For the scope of this study, the data needs to pass all vetoes for CBC searches, meaning that only recording segments with data quality at least \hltt{CBC_CAT3} (using the GWOSC definitions) are used.
            \item \textbf{No hardware injections:} The data on GWOSC does already contain a small number of simulated transient signals called \emph{hardware injections}~\cite{Biwer_2017}.
            We exclude all segments containing such signals.
            \item \textbf{No real signals:} We also exclude the \emph{real} events in \OR1 (\ie GW150914, GW151012, GW151226).
        \end{itemize}

        \subsection{Generating a data set}
        \label{subsec:generating-a-data-set}

        \begin{figure*}
            \centering
            \begin{tikzpicture}

                \tikzstyle{decision} = [diamond, draw, text width=1.75cm, minimum height=1cm, text centered, thick]
                \tikzstyle{block} = [rectangle, draw, text width=5cm, minimum height=1cm, text centered, rounded corners, thick]
                \tikzstyle{wideblock} = [rectangle, draw, text width=7cm, minimum height=1cm, text centered, rounded corners, thick]
                \tikzstyle{line} = [draw, thick, ->]
                \tikzstyle{dotty} = [line, gray!80, line cap=round, thick]
                \tikzset{between/.style args={#1 and #2}{at = ($(#1)!0.5!(#2)$)}}

                \node [wideblock] (config)
                    {Define parameters (and distributions) for waveform simulation in a configuration file};
                \node [wideblock, below = 0.35cm of config] (generate_params)
                    {Randomly sample waveform parameters from given distributions};
                \node [block, below = 0.35cm of config, left = 2cm of generate_params]
                    (generate_time) {Generate random GPS time for noise};
                \node [wideblock, below = 0.35cm of generate_params] (simulate)
                    {Simulate waveform with the given parameters (using PyCBC / LAL)};
                \node [decision, below = 0.35cm of simulate] (success)
                    {Simulation succeeds?};
                \node [draw, between = generate_time.east and generate_params.west, above=2.11cm, thick, rounded corners, inner sep=2mm] (start) 
                    {Start};
                \node [wideblock, below = 0.7cm of success] (fade_on)
                    {Use a one-sided Tukey window to \enquote{fade on} the simulated waveform};
                \node [wideblock, below = 0.35cm of fade_on] (antenna_patterns)
                    {Project waveform onto antenna patterns to make detector signals};
                \node [wideblock, below = 0.35cm of antenna_patterns] (dummy_inject)
                    {Inject the raw simulated waveform into the selected background noise};
                \node [wideblock, below = 0.35cm of dummy_inject] (get_nomf_snr)
                    {Compute the network optimal matched filtering SNR (\texttt{nomf\_snr})};
                \node [wideblock, below = 0.35cm of get_nomf_snr] (scale_factor)
                    {Scale waveform by a factor:\\ \mbox{\texttt{injection\_snr} $\div$ \texttt{nomf\_snr}}};
                \node [wideblock, below = 0.35cm of scale_factor] (add_rescaled)
                    {Inject scaled waveform into the noise};
                \node [wideblock, below = 0.35cm of add_rescaled] (whiten_bandpass)
                    {Whiten and band-pass,\\ crop to specified length};
                \node [wideblock, below = 0.35cm of whiten_bandpass] (save)
                    {Save sample and simulation parameters to an HDF file};
                \node [decision, below = 0.35cm of generate_time] (valid_time)
                    {Is valid noise time?};
                \node [block, below = 0.75cm of valid_time] (select_noise)
                    {Select interval around noise time from LIGO recordings};
                \node [block, left = 2cm of dummy_inject] (downsample)
                    {Downsample noise to \texttt{target\_sampling\_rate}};

                \draw [dotty] (generate_params.west) to["" '] + (-1.25, 0) |- node [near end, left=0.75cm] {Injection SNR} (scale_factor.west);
                \draw [dotty] (config.west) to["" '] + (-0.75, 0) |- node [near end, left=0.5cm, align=right] {Band-pass frequencies,\\ Sample length} (whiten_bandpass.west);
                \draw [line] (config.south) -- (generate_params.north);
                \draw [line] (generate_params.south) -- (simulate.north);
                \draw [line] (simulate.south) -- (success.north);
                \draw [line] (success.east) to["No\\ \small{(e.g., because}\\ \small{of numerical}\\ \small{instabilities)}", align=center, '] + (2.5, 0) |-  (generate_params.east);
                \draw [line] (success.south) to["Yes", '] + (0, -0.5) -- (fade_on);
                \draw [line] (generate_time) -- (valid_time.north);
                \draw [line] (valid_time.west) to["No", '] + (-1.5, 0) |- (generate_time.west);
                \draw [line] (valid_time.south) to["Yes", '] +(0, -0.5) -- (select_noise.north);
                \draw [line] (select_noise.south) -- (downsample.north);
                \draw [line] (fade_on.south) -- (antenna_patterns.north);
                \draw [line] (antenna_patterns.south) -- (dummy_inject.north);
                \draw [line] (dummy_inject.south) -- (get_nomf_snr.north);
                \draw [line] (get_nomf_snr.south) -- (scale_factor.north);
                \draw [line] (scale_factor.south) -- (add_rescaled.north);
                \draw [line] (add_rescaled.south) -- (whiten_bandpass.north);
                \draw [line] (whiten_bandpass.south) -- (save.north);
                \draw [line] (downsample.east) -- (dummy_inject.west);
                \draw [line] (downsample.south) |- (add_rescaled.west);
                \draw [line] (start.west) -| (generate_time.north);
                \draw [line] (start.east) -| (config.north);

            \end{tikzpicture}
            \caption{This flowchart visualizes the process that was used to generate synthetic training and testing data by injecting simulated waveforms into background noise comprised of real LIGO recordings.}
            \label{fig:sample-generation}
        \end{figure*}
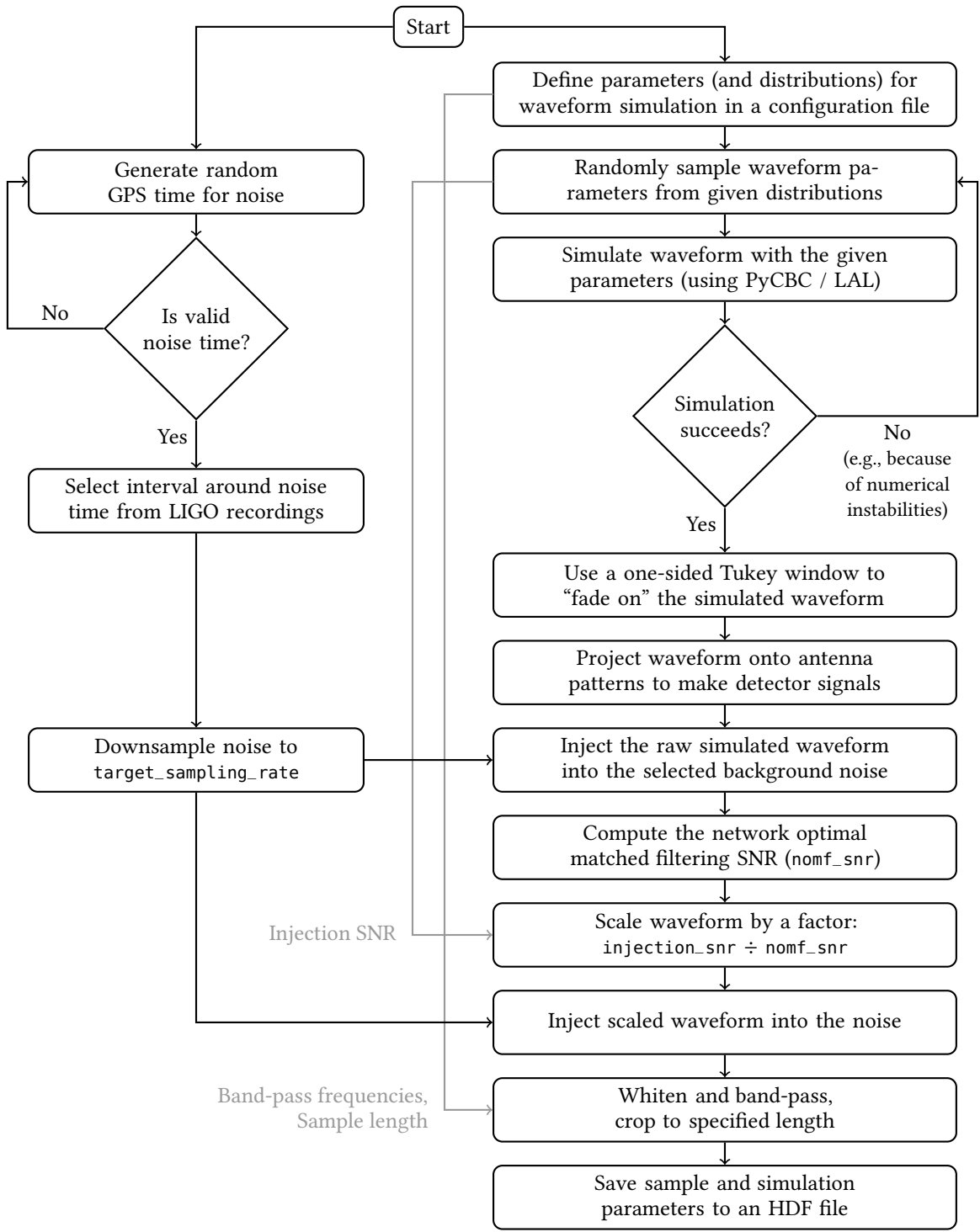

        In this section, we give a detailed account of our data generation process, which is visualized in \cref{fig:sample-generation}.

        In order to generate a new example, we first need to select a piece of LIGO recording to be used as background.
        To this end, we keep drawing a GPS time $t_\text{GPS}$ between the start and end of \OR1 uniformly at random until we find a valid time. 
        A time $t_\text{GPS}$ is considered valid when the symmetric $\delta t$ interval around it fulfills the four criteria defined above.
        To save memory, this interval is then down-sampled from the original sampling  rate of \SI{4096}{\hertz} of the GWOSC data to \SI{2048}{\hertz}.
        Note that $\delta t$ should be chosen larger than half the desired sample length, because we will later compute the (discrete) Fourier transform as part of a whitening procedure. 
        This corrupts the edges at both ends, which need to be cropped off.

        In parallel, a set of parameters for the waveform simulation is sampled from the joint distribution over the entire parameter space.
        This study is limited to waveforms from mergers of binary black holes, which are simulated using the effective-one-body model \hltt{SEOBNRv4} in the time-domain \cite{Bohe_2017}.
        Therefore, we need to randomly sample values for the masses of the black holes, the $z$-components of their spins, the right ascension, declination, polarization, inclination, and coalescence phase angle (which together specify the location and orientation of the source in the sky), as well as the injection SNR.
        For more details about these parameters, see \cref{sec:data-generation-parameters}.

        Choosing the distributions of these parameters is a good example for the contradicting requirements of correctly modeling physics on the one hand and the practical concerns of the ML side.
        In reality, most of the GW signals are expected to be very faint, because their sources are comparatively far away:
        If we assume the sources to be distributed isotropically and uniformly in space over the whole (spherical) search volume, approximately half of all sources will be at 80\% or more of the maximum sensitive distance.
        However, if this $r^3$-dependency is modeled correctly when sampling parameters for simulating training data, a large fraction of the data will be barely above the detectability threshold.
        This makes it hard for the machine learning methods to actually learn anything.
        One common approach in deep learning to address this kind of problem is to split the training into different phases, first training on \enquote{easy} examples (in this case events with strong GW signals), and then gradually replacing or complementing the training set with \enquote{harder} (\ie fainter) examples.
        In our experiments, however, this so-called \emph{curriculum learning} \cite{Bengio_2009} did not lead to relevant improvements of the final performance.

        The simulation routines in \textsc{LALSuite} return two time series for given parameter settings, namely the two polarization modes of the gravitational wave, $h_+$ and $h_\cross$.
        These are then transformed according to the interferometer \emph{antenna patterns}, which are functions that describe the directional sensitivity of the detector \footnote{For further details, see, for example, section 9.5.3 (b) in \cite{Thorne_1987}, section 9.2.3 in \cite{Maggiore_2008}, or section 6.1.11 in \cite{Creighton_2011}.}.
        \pycbc{} provides methods to calculate the projection onto the antenna patterns  for the detectors in Hanford and Livingston for a given source location in the sky and a corresponding polarization angle.
        Finally, the simulated detector signals also need to be corrected for the time offset between the detectors, based again on the relative source location in the sky.
        This gives us the \enquote{pure} signals that the detectors would observe in the absence of noise.

        Next, these signals are \emph{injected} into the noise that we selected in the beginning.
        For comparison later on, we would like to know how \enquote{loud} the injection was.
        This can be measured by the \emph{optimal matched filter SNR} (\eg \cite{Cutler_1994, Smith_2013}) of the injection, which is the maximal SNR possible resulting from using the time-inverted signal itself as a filter.
        This is achieved by a two-step process:
        \begin{enumerate}
            \item First, we simply add the two time series (noise and signal) in such a fashion that the peak of the signal amplitude in H1 is centered within the noise interval.
            Afterwards, we compute the optimal matched filtering signal-to-noise ratio in both detectors, and subsequently also the network optimal matched filtering SNR (NOMF-SNR).
            The latter is then used to determine a scaling factor by which the waveform needs to be multiplied to ensure that the injected signal has the desired \emph{injection SNR}.
            This is possible because multiplying the waveforms of \emph{both} detectors by a factor $\lambda$ results in a network SNR that has been scaled by the same factor $\lambda$.
            From an astrophysical perspective, rescaling simply corresponds to moving the source closer or further away from the detectors.
            \item Now we can add the rescaled waveform to the noise, which guarantees that the sample has the desired network SNR.
        \end{enumerate}

        The result is then whitened with \pycbc{} using a local estimate of the power spectral density, and high-passed at \SI{20}{\hertz} to remove some of the non-physical turn-on artifacts from the simulation.
        Finally, the example is cropped to the desired length (which was chosen as \SI{8}{\second}) in such a fashion that the maximum of the signal always ends up at the same (relative) location within the sample.
        This is permitted, because our particular choice of model architecture (see below) is not sensitive to the position of the signal within a sample.
        The choice of \SI{8}{\second} for the length was governed by memory considerations: training a neural network efficiently requires that both a minibatch of training examples and the network parameters (together with their gradients) fit into memory of a graphical processing unit (GPU).

        \subsection{Training and testing datasets}
        \label{subsec:training-and-testing-data-sets}

        For this work, we created three datasets:
        a training dataset with \num{32768} examples, a validation set with \num{4096} examples, and a testing dataset with \num{16384} examples.
        The parameters for the waveform simulation were drawn independently from the same joint distribution over the parameter space (see \cref{sec:data-generation-parameters}) for all three data sets. 
        All data sets are mutually disjoint, that is, no single example is used for both training and testing~/~validation.

        To ensure that during training the net is also exposed to sufficient data which do not contain any signals, a number of examples not containing any injections is generated by simply skipping the injection step.
        We use three times as many examples that contain an injection than pure noise examples.

        In \cref{sec:experiments-and-results}, we also evaluate our trained model on real signals from LIGO's first observation run, which have undergone analogous preprocessing (whitening, band-passing) like the training data.

        \section{Model and training procedure}
        \label{sec:model-and-training-procedure}

\begin{figure*}[t]
            \centering
            \begin{tikzpicture}

                \tikzstyle{box} = [draw=black, fill=white, thick, rectangle, anchor=west, rounded corners, minimum height=0.7cm]
                \tikzstyle{circ} = [circle, minimum height=2.85mm, inner sep=0mm]
                \tikzstyle{cfill} = [draw=color1, very thick, fill=white]
                \tikzstyle{cdot} = [circle, fill=color1, minimum size=0.75mm, inner sep=0mm]
                \tikzstyle{dashy} = [thick, dashed, line cap=round, gray!50]
                \tikzstyle{dilation} = [anchor=center, align=center, font=\scriptsize, inner sep=0.25mm, fill=white]
                \tikzstyle{arr} = [very thick, line cap=round, postaction={decorate}, decoration={markings, mark=at position 0.55 with {\arrow{>}}}]

                \foreach \y in {3} { \foreach \x in {3.5,4.0,...,8.5} { \draw [dashy] (\x, \y-0.65) -- (\x, \y); } \foreach \x in {9.5,10.0,...,14.5} { \draw [dashy] (\x, \y-0.65) -- (\x, \y); } }
                \foreach \y in {4} { \foreach \x in {4.0,4.5,...,8.5} { \draw [dashy] (\x, \y) -- (\x, \y-1); \draw [dashy] (\x, \y) -- (\x-0.5, \y-1); } \foreach \x in {10.0,10.5,...,14.5} { \draw [dashy] (\x, \y) -- (\x, \y-1); \draw [dashy] (\x, \y) -- (\x-0.5, \y-1); } }
                \foreach \y in {5} { \foreach \x in {4.5,5.0,...,8.0} { \draw [dashy] (\x, \y) -- (\x-0.5, \y-1); \draw [dashy] (\x, \y) -- (\x+0.5, \y-1); } \foreach \x in {10.0,10.5,...,14.0} { \draw [dashy] (\x, \y) -- (\x-0.5, \y-1); \draw [dashy] (\x, \y) -- (\x+0.5, \y-1); } }
                \foreach \y in {6} { \foreach \x in {5.5,6.0,...,7.5} { \draw [dashy] (\x, \y) -- (\x-1.0, \y-1); \draw [dashy] (\x, \y) -- (\x+1.0, \y-1); } \foreach \x in {10.5,11.0,...,13.0} { \draw [dashy] (\x, \y) -- (\x-1.0, \y-1); \draw [dashy] (\x, \y) -- (\x+1.0, \y-1); } }

                \node [circ, fill] (1-6) at (6.5, 6) {};
                \node [circ] (1-5) at (5.5, 5) {};
                \node [circ] (2-5) at (7.5, 5) {};
                \node [circ] (1-4) at (5.0, 4) {};
                \node [circ] (2-4) at (6.0, 4) {};
                \node [circ] (3-4) at (7.0, 4) {};
                \node [circ] (4-4) at (8.0, 4) {};
                \node [circ] (1-3) at (4.5, 3) {};
                \node [circ] (2-3) at (5.0, 3) {};
                \node [circ] (3-3) at (5.5, 3) {};
                \node [circ] (4-3) at (6.0, 3) {};
                \node [circ] (5-3) at (6.5, 3) {};
                \node [circ] (6-3) at (7.0, 3) {};
                \node [circ] (7-3) at (7.5, 3) {};
                \node [circ] (8-3) at (8.0, 3) {};

                \draw [arr] (1-5) -- (1-6);
                \draw [arr] (2-5) -- (1-6);
                \draw [arr] (1-4) -- (1-5);
                \draw [arr] (2-4) -- (1-5);
                \draw [arr] (3-4) -- (2-5);
                \draw [arr] (4-4) -- (2-5);
                \draw [arr] (1-3) -- (1-4);
                \draw [arr] (2-3) -- (1-4);
                \draw [arr] (3-3) -- (2-4);
                \draw [arr] (4-3) -- (2-4);
                \draw [arr] (5-3) -- (3-4);
                \draw [arr] (6-3) -- (3-4);
                \draw [arr] (7-3) -- (4-4);
                \draw [arr] (8-3) -- (4-4);
                \draw [arr] (1-3) ++(0, -0.65) -- (1-3);
                \draw [arr] (2-3) ++(0, -0.65) -- (2-3);
                \draw [arr] (3-3) ++(0, -0.65) -- (3-3);
                \draw [arr] (4-3) ++(0, -0.65) -- (4-3);
                \draw [arr] (5-3) ++(0, -0.65) -- (5-3);
                \draw [arr] (6-3) ++(0, -0.65) -- (6-3);
                \draw [arr] (7-3) ++(0, -0.65) -- (7-3);
                \draw [arr] (8-3) ++(0, -0.65) -- (8-3);

                \foreach \y in {3} { \foreach \x in {3.5,4.0,...,8.5} { \node [circ, cfill] at (\x, \y) {}; } \foreach \x in {9.5,10.0,...,14.5} { \node [circ, cfill] at (\x, \y) {}; } }
                \foreach \y in {4} { \foreach \x in {4.0,4.5,...,8.5} { \node [circ, cfill] at (\x, \y) {}; } \foreach \x in {9.5,10.0,...,14.5} { \node [circ, cfill] at (\x, \y) {}; } }
                \foreach \y in {5} { \foreach \x in {4.5,5.0,...,8.5} { \node [circ, cfill] at (\x, \y) {}; } \foreach \x in {9.5,10.0,...,14.0} { \node [circ, cfill] at (\x, \y) {}; } }
                \foreach \y in {6} { \foreach \x in {5.5,6.0,...,8.5} { \node [circ, cfill] at (\x, \y) {}; } \foreach \x in {9.5,10.0,...,13.0} { \node [circ, cfill] at (\x, \y) {}; } }
                \node [circ, fill=color1] at (6.5, 6) {};
                \foreach \y in {3, 4, 5, 6} { \foreach \x in {9.0} { \node [cdot, xshift=-1.5mm] at (\x, \y) {}; \node [cdot, xshift= 0.0mm] at (\x, \y) {}; \node [cdot, xshift= 1.5mm] at (\x, \y) {}; } }
                \foreach \y in {6.5} { \foreach \x in {5.5,6.0,...,8.5} { \node [cdot, yshift=-1.5mm] at (\x, \y) {}; \node [cdot, yshift= 0.0mm] at (\x, \y) {}; \node [cdot, yshift= 1.5mm] at (\x, \y) {}; } \foreach \x in {9.5,10.0,...,13.0} { \node [cdot, yshift=-1.5mm] at (\x, \y) {}; \node [cdot, yshift= 0.0mm] at (\x, \y) {}; \node [cdot, yshift= 1.5mm] at (\x, \y) {}; } }
                \foreach \y in {1.30} { \foreach \x in {3.5,4.0,...,14.5} { \draw [very thick, gray!50, ->] (\x, \y-0.2) -- (\x, \y+0.2); } }
                \foreach \y in {8.90} { \foreach \x in {5.5,6.0,...,13.0} { \draw [very thick, gray!50, ->] (\x, \y-0.2) -- (\x, \y+0.2); } }
                \foreach \y in {7.95} { \foreach \x in {5.5,6.0,...,13.0} { \draw [gray!50, very thick] (\x, \y-0.4) -- (\x, \y); } }
                \node [box, minimum width=12cm] at (3.0, 2.0) 
                    {2 Channels $\to$ 512 Channels};
                \node [box, minimum width=8.5cm] at (5.0, 7.25) 
                    {512 Channels $\to$ 1 Channel};
                \node [box, minimum width=8.5cm] at (5.0, 8.25) 
                    {$\sigma(x):\:\mathbb{R} \to [0, 1]$};
                \node [anchor=center, align=center, inner sep=0.75mm] at (1.5, 0.75) (label_1)
                    {Strain H1/L1};
                \node [anchor=center, align=center, inner sep=0.75mm] at (1.5, 2.00) (label_2)
                    {Input\\ ConvLayer};
                \node [anchor=center, align=center, inner sep=0.75mm] at (1.5, 4.75) (label_3)
                    {Stack of\\ 12 blocks\\ with dilated\\ ConvLayers};
                \node [anchor=center, align=center, inner sep=0.75mm] at (1.5, 7.25) (label_4)
                    {Output\\ ConvLayer};
                \node [anchor=center, align=center, inner sep=0.75mm] at (1.5, 8.25) (label_5)
                    {Sigmoid Layer};
                \node [anchor=center, align=center, inner sep=0.75mm] at (1.5, 9.25) (label_6)
                    {(Raw) Prediction};
                \node [dilation] at (9.0, 3.50)
                    {Dilation 1};
                \node [dilation] at (9.0, 4.50)
                    {Dilation 2};
                \node [dilation] at (9.0, 5.50)
                    {Dilation 4};

                \draw [very thick, ->] (label_1.north) -- (label_2.south);
                \draw [very thick, ->] (label_2.north) -- (label_3.south);
                \draw [very thick, ->] (label_3.north) -- (label_4.south);
                \draw [very thick, ->] (label_4.north) -- (label_5.south);
                \draw [very thick, ->] (label_5.north) -- (label_6.south);

                \draw [color2, line width=0.4mm] plot [smooth] coordinates{ (3.000, +0.773) (3.020, +0.579) (3.040, +0.754) (3.060, +0.814) (3.080, +0.655) (3.100, +0.900) (3.120, +0.791) (3.140, +0.631) (3.160, +0.835) (3.180, +0.706) (3.200, +0.720) (3.220, +0.966) (3.240, +0.746) (3.260, +0.774) (3.280, +0.771) (3.300, +0.671) (3.320, +0.860) (3.340, +0.591) (3.360, +0.688) (3.380, +0.881) (3.400, +0.892) (3.420, +0.865) (3.440, +0.699) (3.460, +0.752) (3.480, +0.883) (3.500, +0.722) (3.520, +0.692) (3.540, +0.754) (3.560, +0.842) (3.580, +0.765) (3.600, +0.870) (3.620, +0.713) (3.640, +0.795) (3.660, +0.810) (3.680, +0.816) (3.700, +0.725) (3.720, +0.845) (3.740, +0.727) (3.760, +0.787) (3.780, +0.643) (3.800, +0.767) (3.820, +0.785) (3.840, +0.605) (3.860, +0.735) (3.880, +0.693) (3.900, +0.699) (3.920, +0.704) (3.940, +0.758) (3.960, +0.814) (3.980, +0.856) (4.000, +0.745) (4.020, +0.764) (4.040, +0.900) (4.060, +0.664) (4.080, +0.760) (4.100, +0.815) (4.120, +0.818) (4.140, +0.736) (4.160, +0.637) (4.180, +0.724) (4.200, +0.624) (4.220, +0.843) (4.240, +0.822) (4.260, +0.848) (4.280, +0.806) (4.300, +0.792) (4.320, +0.732) (4.340, +0.840) (4.360, +0.751) (4.380, +0.849) (4.400, +0.620) (4.420, +0.639) (4.440, +0.851) (4.460, +0.726) (4.480, +0.757) (4.500, +0.736) (4.520, +0.792) (4.540, +0.696) (4.560, +0.546) (4.580, +0.660) (4.600, +0.906) (4.620, +0.891) (4.640, +0.639) (4.660, +0.668) (4.680, +0.727) (4.700, +0.814) (4.720, +0.739) (4.740, +0.652) (4.760, +0.821) (4.780, +0.735) (4.800, +0.618) (4.820, +0.553) (4.840, +0.772) (4.860, +0.658) (4.880, +0.781) (4.900, +0.884) (4.920, +0.779) (4.940, +0.830) (4.960, +0.640) (4.980, +0.755) (5.000, +0.840) (5.020, +0.907) (5.040, +0.800) (5.060, +0.798) (5.080, +0.560) (5.100, +0.725) (5.120, +0.771) (5.140, +0.812) (5.160, +0.512) (5.180, +0.700) (5.200, +0.597) (5.220, +0.522) (5.240, +0.568) (5.260, +0.826) (5.280, +0.857) (5.300, +0.626) (5.320, +0.593) (5.340, +0.542) (5.360, +0.764) (5.380, +0.806) (5.400, +0.683) (5.420, +0.804) (5.440, +0.816) (5.460, +0.773) (5.480, +0.718) (5.500, +0.746) (5.520, +0.766) (5.540, +0.862) (5.560, +1.041) (5.580, +0.830) (5.600, +0.818) (5.620, +0.694) (5.640, +0.789) (5.660, +0.944) (5.680, +0.782) (5.700, +0.643) (5.720, +0.872) (5.740, +0.780) (5.760, +0.851) (5.780, +0.711) (5.800, +0.799) (5.820, +0.596) (5.840, +0.750) (5.860, +0.788) (5.880, +0.789) (5.900, +0.825) (5.920, +0.701) (5.940, +0.736) (5.960, +0.810) (5.980, +0.660) (6.000, +0.656) (6.020, +0.860) (6.040, +0.770) (6.060, +0.622) (6.080, +0.671) (6.100, +0.722) (6.120, +0.601) (6.140, +0.817) (6.160, +0.853) (6.180, +0.793) (6.200, +0.829) (6.220, +0.824) (6.240, +0.789) (6.260, +0.699) (6.280, +0.667) (6.300, +0.818) (6.320, +0.673) (6.340, +0.840) (6.360, +0.923) (6.380, +0.709) (6.400, +0.873) (6.420, +0.855) (6.440, +0.731) (6.460, +0.797) (6.480, +0.761) (6.500, +0.610) (6.520, +0.843) (6.540, +0.735) (6.560, +0.774) (6.580, +0.651) (6.600, +0.761) (6.620, +0.807) (6.640, +0.743) (6.660, +0.825) (6.680, +0.753) (6.700, +0.782) (6.720, +0.708) (6.740, +0.740) (6.760, +0.710) (6.780, +0.707) (6.800, +0.700) (6.820, +0.607) (6.840, +0.792) (6.860, +0.682) (6.880, +0.775) (6.900, +0.810) (6.920, +0.745) (6.940, +0.745) (6.960, +0.685) (6.980, +0.782) (7.000, +0.798) (7.020, +0.755) (7.040, +0.869) (7.060, +0.781) (7.080, +0.776) (7.100, +0.738) (7.120, +0.731) (7.140, +0.803) (7.160, +0.819) (7.180, +0.678) (7.200, +0.871) (7.220, +0.857) (7.240, +0.774) (7.260, +0.741) (7.280, +0.756) (7.300, +0.839) (7.320, +0.595) (7.340, +0.685) (7.360, +0.622) (7.380, +0.723) (7.400, +0.892) (7.420, +0.798) (7.440, +0.721) (7.460, +0.582) (7.480, +0.792) (7.500, +0.707) (7.520, +0.745) (7.540, +0.720) (7.560, +0.539) (7.580, +0.549) (7.600, +0.757) (7.620, +0.820) (7.640, +0.743) (7.660, +0.769) (7.680, +0.693) (7.700, +0.708) (7.720, +0.696) (7.740, +0.760) (7.760, +0.920) (7.780, +0.745) (7.800, +0.689) (7.820, +0.788) (7.840, +0.655) (7.860, +0.811) (7.880, +0.772) (7.900, +0.836) (7.920, +0.736) (7.940, +0.694) (7.960, +0.698) (7.980, +0.645) (8.000, +0.647) (8.020, +0.823) (8.040, +0.675) (8.060, +0.542) (8.080, +0.614) (8.100, +0.667) (8.120, +0.620) (8.140, +0.718) (8.160, +0.795) (8.180, +0.626) (8.200, +0.724) (8.220, +0.722) (8.240, +0.763) (8.260, +0.807) (8.280, +0.807) (8.300, +0.657) (8.320, +0.739) (8.340, +0.779) (8.360, +0.619) (8.380, +0.729) (8.400, +0.704) (8.420, +0.848) (8.440, +0.766) (8.460, +0.715) (8.480, +0.666) (8.500, +0.749) (8.520, +0.899) (8.540, +0.789) (8.560, +0.787) (8.580, +0.601) (8.600, +0.700) (8.620, +0.756) (8.640, +0.825) (8.660, +0.627) (8.680, +0.847) (8.700, +0.744) (8.720, +0.945) (8.740, +0.870) (8.760, +0.728) (8.780, +0.808) (8.800, +0.775) (8.820, +0.910) (8.840, +0.686) (8.860, +0.910) (8.880, +0.776) (8.900, +0.690) (8.920, +0.734) (8.940, +0.726) (8.960, +0.707) (8.980, +0.827) (9.000, +0.748) (9.020, +0.634) (9.040, +0.632) (9.060, +0.668) (9.080, +0.854) (9.100, +0.886) (9.120, +0.759) (9.140, +0.629) (9.160, +0.856) (9.180, +0.834) (9.200, +0.657) (9.220, +0.766) (9.240, +0.744) (9.260, +0.884) (9.280, +0.883) (9.300, +0.739) (9.320, +0.714) (9.340, +0.746) (9.360, +0.727) (9.380, +0.783) (9.400, +0.578) (9.420, +0.924) (9.440, +0.743) (9.460, +0.803) (9.480, +0.774) (9.500, +0.732) (9.520, +0.677) (9.540, +0.675) (9.560, +0.754) (9.580, +0.831) (9.600, +0.683) (9.620, +0.676) (9.640, +0.731) (9.660, +0.936) (9.680, +0.809) (9.700, +0.709) (9.720, +0.826) (9.740, +0.718) (9.760, +0.744) (9.780, +0.761) (9.800, +0.736) (9.820, +0.804) (9.840, +0.756) (9.860, +0.729) (9.880, +0.779) (9.900, +0.781) (9.920, +0.843) (9.940, +0.757) (9.960, +0.752) (9.980, +0.724) (10.000, +0.789) (10.020, +0.790) (10.040, +0.760) (10.060, +0.842) (10.080, +0.675) (10.100, +0.700) (10.120, +0.771) (10.140, +0.783) (10.160, +0.640) (10.180, +0.719) (10.200, +0.696) (10.220, +0.645) (10.240, +0.691) (10.260, +0.787) (10.280, +0.784) (10.300, +0.782) (10.320, +0.832) (10.340, +0.846) (10.360, +0.666) (10.380, +0.842) (10.400, +0.731) (10.420, +0.832) (10.440, +0.708) (10.460, +0.843) (10.480, +0.726) (10.500, +0.651) (10.520, +0.739) (10.540, +0.840) (10.560, +0.834) (10.580, +0.777) (10.600, +0.731) (10.620, +0.701) (10.640, +0.831) (10.660, +0.898) (10.680, +0.746) (10.700, +0.834) (10.720, +0.809) (10.740, +0.642) (10.760, +0.746) (10.780, +0.676) (10.800, +0.847) (10.820, +0.640) (10.840, +0.813) (10.860, +0.777) (10.880, +0.811) (10.900, +0.579) (10.920, +0.789) (10.940, +0.758) (10.960, +0.741) (10.980, +0.924) (11.000, +0.584) (11.020, +0.753) (11.040, +0.641) (11.060, +0.721) (11.080, +0.727) (11.100, +0.833) (11.120, +0.835) (11.140, +0.715) (11.160, +0.747) (11.180, +0.682) (11.200, +0.583) (11.220, +0.688) (11.240, +0.657) (11.260, +0.815) (11.280, +0.697) (11.300, +0.901) (11.320, +0.813) (11.340, +0.864) (11.360, +0.759) (11.380, +0.686) (11.400, +0.649) (11.420, +0.701) (11.440, +0.700) (11.460, +0.700) (11.480, +0.911) (11.500, +0.853) (11.520, +0.938) (11.540, +0.802) (11.560, +0.923) (11.580, +0.630) (11.600, +0.539) (11.620, +0.588) (11.640, +0.454) (11.660, +0.544) (11.680, +0.656) (11.700, +0.605) (11.720, +1.098) (11.740, +0.912) (11.760, +1.150) (11.780, +1.022) (11.800, +0.892) (11.820, +0.869) (11.840, +0.709) (11.860, +0.475) (11.880, +0.451) (11.900, +0.408) (11.920, +0.533) (11.940, +0.775) (11.960, +0.940) (11.980, +1.056) (12.000, +0.723) (12.020, +0.713) (12.040, +0.707) (12.060, +0.798) (12.080, +0.783) (12.100, +0.744) (12.120, +0.909) (12.140, +0.758) (12.160, +0.862) (12.180, +0.737) (12.200, +0.760) (12.220, +0.840) (12.240, +0.817) (12.260, +0.772) (12.280, +0.771) (12.300, +0.794) (12.320, +0.705) (12.340, +0.728) (12.360, +0.573) (12.380, +0.763) (12.400, +0.854) (12.420, +0.834) (12.440, +0.753) (12.460, +0.796) (12.480, +0.851) (12.500, +0.740) (12.520, +0.722) (12.540, +0.907) (12.560, +0.789) (12.580, +0.782) (12.600, +0.701) (12.620, +0.749) (12.640, +0.751) (12.660, +0.493) (12.680, +0.680) (12.700, +0.691) (12.720, +0.749) (12.740, +0.776) (12.760, +0.668) (12.780, +0.749) (12.800, +0.770) (12.820, +0.820) (12.840, +0.819) (12.860, +0.670) (12.880, +0.683) (12.900, +0.662) (12.920, +0.667) (12.940, +0.783) (12.960, +0.778) (12.980, +0.738) (13.000, +0.622) (13.020, +0.708) (13.040, +0.763) (13.060, +0.758) (13.080, +0.786) (13.100, +0.571) (13.120, +0.721) (13.140, +0.773) (13.160, +0.546) (13.180, +0.756) (13.200, +0.867) (13.220, +0.963) (13.240, +0.837) (13.260, +0.803) (13.280, +0.772) (13.300, +0.658) (13.320, +0.677) (13.340, +0.686) (13.360, +0.710) (13.380, +0.638) (13.400, +0.672) (13.420, +0.643) (13.440, +0.678) (13.460, +0.726) (13.480, +0.788) (13.500, +0.766) (13.520, +0.713) (13.540, +0.736) (13.560, +0.708) (13.580, +0.749) (13.600, +0.813) (13.620, +0.842) (13.640, +0.693) (13.660, +0.660) (13.680, +0.650) (13.700, +0.661) (13.720, +0.698) (13.740, +0.649) (13.760, +0.807) (13.780, +0.767) (13.800, +0.763) (13.820, +0.806) (13.840, +0.696) (13.860, +0.742) (13.880, +0.819) (13.900, +0.702) (13.920, +0.644) (13.940, +0.599) (13.960, +0.803) (13.980, +0.817) (14.000, +0.726) (14.020, +0.595) (14.040, +0.659) (14.060, +0.966) (14.080, +0.758) (14.100, +0.720) (14.120, +0.765) (14.140, +0.804) (14.160, +0.823) (14.180, +0.511) (14.200, +0.793) (14.220, +0.798) (14.240, +0.761) (14.260, +0.745) (14.280, +0.715) (14.300, +0.722) (14.320, +0.757) (14.340, +0.775) (14.360, +0.709) (14.380, +0.867) (14.400, +0.826) (14.420, +0.773) (14.440, +0.883) (14.460, +0.771) (14.480, +0.743) (14.500, +0.690) (14.520, +0.849) (14.540, +0.641) (14.560, +0.764) (14.580, +0.741) (14.600, +0.794) (14.620, +0.783) (14.640, +0.715) (14.660, +0.826) (14.680, +0.732) (14.700, +0.876) (14.720, +0.778) (14.740, +0.854) (14.760, +0.695) (14.780, +0.708) (14.800, +0.678) (14.820, +0.645) (14.840, +0.859) (14.860, +0.731) (14.880, +1.004) (14.900, +0.719) (14.920, +0.767) (14.940, +0.746) (14.960, +0.722) (14.980, +0.810) };
                \draw [densely dotted, gray] (3, 0.75) -- (15, 0.75);
                \draw [color0, line width=0.45mm] plot [smooth] coordinates{ (5.000, +9.270) (5.020, +9.270) (5.040, +9.274) (5.060, +9.266) (5.080, +9.281) (5.100, +9.279) (5.120, +9.281) (5.140, +9.275) (5.160, +9.275) (5.180, +9.266) (5.200, +9.279) (5.220, +9.275) (5.240, +9.275) (5.260, +9.255) (5.280, +9.263) (5.300, +9.271) (5.320, +9.285) (5.340, +9.283) (5.360, +9.285) (5.380, +9.284) (5.400, +9.289) (5.420, +9.281) (5.440, +9.267) (5.460, +9.269) (5.480, +9.266) (5.500, +9.272) (5.520, +9.259) (5.540, +9.265) (5.560, +9.261) (5.580, +9.273) (5.600, +9.267) (5.620, +9.270) (5.640, +9.264) (5.660, +9.267) (5.680, +9.265) (5.700, +9.269) (5.720, +9.264) (5.740, +9.266) (5.760, +9.267) (5.780, +9.266) (5.800, +9.266) (5.820, +9.270) (5.840, +9.270) (5.860, +9.272) (5.880, +9.262) (5.900, +9.266) (5.920, +9.261) (5.940, +9.262) (5.960, +9.268) (5.980, +9.268) (6.000, +9.263) (6.020, +9.267) (6.040, +9.275) (6.060, +9.275) (6.080, +9.266) (6.100, +9.264) (6.120, +9.265) (6.140, +9.270) (6.160, +9.265) (6.180, +9.277) (6.200, +9.275) (6.220, +9.279) (6.240, +9.277) (6.260, +9.273) (6.280, +9.270) (6.300, +9.262) (6.320, +9.265) (6.340, +9.261) (6.360, +9.269) (6.380, +9.273) (6.400, +9.284) (6.420, +9.285) (6.440, +9.274) (6.460, +9.263) (6.480, +9.254) (6.500, +9.256) (6.520, +9.261) (6.540, +9.280) (6.560, +9.285) (6.580, +9.295) (6.600, +9.289) (6.620, +9.291) (6.640, +9.284) (6.660, +9.272) (6.680, +9.267) (6.700, +9.260) (6.720, +9.267) (6.740, +9.268) (6.760, +9.269) (6.780, +9.272) (6.800, +9.285) (6.820, +9.285) (6.840, +9.281) (6.860, +9.265) (6.880, +9.276) (6.900, +9.270) (6.920, +9.274) (6.940, +9.272) (6.960, +9.270) (6.980, +9.271) (7.000, +9.275) (7.020, +9.280) (7.040, +9.276) (7.060, +9.279) (7.080, +9.277) (7.100, +9.274) (7.120, +9.261) (7.140, +9.282) (7.160, +9.285) (7.180, +9.294) (7.200, +9.293) (7.220, +9.307) (7.240, +9.299) (7.260, +9.287) (7.280, +9.281) (7.300, +9.289) (7.320, +9.299) (7.340, +9.288) (7.360, +9.278) (7.380, +9.264) (7.400, +9.268) (7.420, +9.269) (7.440, +9.264) (7.460, +9.262) (7.480, +9.256) (7.500, +9.255) (7.520, +9.263) (7.540, +9.292) (7.560, +9.299) (7.580, +9.294) (7.600, +9.271) (7.620, +9.266) (7.640, +9.279) (7.660, +9.277) (7.680, +9.284) (7.700, +9.276) (7.720, +9.276) (7.740, +9.276) (7.760, +9.267) (7.780, +9.269) (7.800, +9.274) (7.820, +9.271) (7.840, +9.269) (7.860, +9.258) (7.880, +9.265) (7.900, +9.266) (7.920, +9.264) (7.940, +9.262) (7.960, +9.267) (7.980, +9.275) (8.000, +9.280) (8.020, +9.273) (8.040, +9.276) (8.060, +9.273) (8.080, +9.274) (8.100, +9.276) (8.120, +9.275) (8.140, +9.282) (8.160, +9.271) (8.180, +9.273) (8.200, +9.270) (8.220, +9.269) (8.240, +9.267) (8.260, +9.267) (8.280, +9.270) (8.300, +9.273) (8.320, +9.274) (8.340, +9.284) (8.360, +9.281) (8.380, +9.284) (8.400, +9.277) (8.420, +9.275) (8.440, +9.267) (8.460, +9.258) (8.480, +9.270) (8.500, +9.275) (8.520, +9.275) (8.540, +9.263) (8.560, +9.264) (8.580, +9.264) (8.600, +9.267) (8.620, +9.258) (8.640, +9.264) (8.660, +9.259) (8.680, +9.261) (8.700, +9.258) (8.720, +9.258) (8.740, +9.259) (8.760, +9.259) (8.780, +9.263) (8.800, +9.274) (8.820, +9.274) (8.840, +9.275) (8.860, +9.264) (8.880, +9.265) (8.900, +9.259) (8.920, +9.257) (8.940, +9.258) (8.960, +9.260) (8.980, +9.265) (9.000, +9.259) (9.020, +9.267) (9.040, +9.266) (9.060, +9.268) (9.080, +9.257) (9.100, +9.256) (9.120, +9.258) (9.140, +9.264) (9.160, +9.269) (9.180, +9.276) (9.200, +9.280) (9.220, +9.275) (9.240, +9.264) (9.260, +9.254) (9.280, +9.261) (9.300, +9.275) (9.320, +9.281) (9.340, +9.285) (9.360, +9.272) (9.380, +9.280) (9.400, +9.272) (9.420, +9.272) (9.440, +9.275) (9.460, +9.274) (9.480, +9.275) (9.500, +9.259) (9.520, +9.258) (9.540, +9.275) (9.560, +9.295) (9.580, +9.292) (9.600, +9.278) (9.620, +9.259) (9.640, +9.260) (9.660, +9.258) (9.680, +9.262) (9.700, +9.265) (9.720, +9.261) (9.740, +9.274) (9.760, +9.269) (9.780, +9.274) (9.800, +9.260) (9.820, +9.270) (9.840, +9.270) (9.860, +9.268) (9.880, +9.267) (9.900, +9.262) (9.920, +9.266) (9.940, +9.262) (9.960, +9.272) (9.980, +9.276) (10.000, +9.278) (10.020, +9.275) (10.040, +9.286) (10.060, +9.292) (10.080, +9.293) (10.100, +9.285) (10.120, +9.275) (10.140, +9.271) (10.160, +9.270) (10.180, +9.270) (10.200, +9.268) (10.220, +9.257) (10.240, +9.260) (10.260, +9.263) (10.280, +9.271) (10.300, +9.266) (10.320, +9.263) (10.340, +9.267) (10.360, +9.268) (10.380, +9.270) (10.400, +9.267) (10.420, +9.266) (10.440, +9.265) (10.460, +9.264) (10.480, +9.262) (10.500, +9.274) (10.520, +9.269) (10.540, +9.273) (10.560, +9.273) (10.580, +9.274) (10.600, +9.271) (10.620, +9.263) (10.640, +9.271) (10.660, +9.280) (10.680, +9.273) (10.700, +9.280) (10.720, +9.282) (10.740, +9.284) (10.760, +9.270) (10.780, +9.261) (10.800, +9.274) (10.820, +9.275) (10.840, +9.289) (10.860, +9.275) (10.880, +9.275) (10.900, +9.260) (10.920, +9.260) (10.940, +9.258) (10.960, +9.265) (10.980, +9.262) (11.000, +9.270) (11.020, +9.274) (11.040, +9.282) (11.060, +9.281) (11.080, +9.282) (11.100, +9.275) (11.120, +9.277) (11.140, +9.274) (11.160, +9.281) (11.180, +9.279) (11.200, +9.269) (11.220, +9.262) (11.240, +9.266) (11.260, +9.278) (11.280, +9.278) (11.300, +9.268) (11.320, +9.255) (11.340, +9.256) (11.360, +9.256) (11.380, +9.273) (11.400, +9.288) (11.420, +9.286) (11.440, +9.274) (11.460, +9.259) (11.480, +9.260) (11.500, +9.262) (11.520, +9.267) (11.540, +9.265) (11.560, +9.266) (11.580, +9.265) (11.600, +9.272) (11.620, +9.266) (11.640, +9.278) (11.660, +9.277) (11.680, +9.440) (11.700, +9.586) (11.720, +9.745) (11.740, +9.745) (11.760, +9.745) (11.780, +9.745) (11.800, +9.745) (11.820, +9.745) (11.840, +9.745) (11.860, +9.745) (11.880, +9.583) (11.900, +9.427) (11.920, +9.263) (11.940, +9.260) (11.960, +9.253) (11.980, +9.257) (12.000, +9.261) (12.020, +9.259) (12.040, +9.264) (12.060, +9.268) (12.080, +9.272) (12.100, +9.265) (12.120, +9.260) (12.140, +9.266) (12.160, +9.267) (12.180, +9.269) (12.200, +9.269) (12.220, +9.272) (12.240, +9.270) (12.260, +9.263) (12.280, +9.260) (12.300, +9.264) (12.320, +9.271) (12.340, +9.276) (12.360, +9.278) (12.380, +9.269) (12.400, +9.270) (12.420, +9.264) (12.440, +9.272) (12.460, +9.266) (12.480, +9.272) (12.500, +9.265) (12.520, +9.271) (12.540, +9.270) (12.560, +9.272) (12.580, +9.264) (12.600, +9.260) (12.620, +9.264) (12.640, +9.276) (12.660, +9.273) (12.680, +9.273) (12.700, +9.265) (12.720, +9.273) (12.740, +9.267) (12.760, +9.267) (12.780, +9.268) (12.800, +9.278) (12.820, +9.277) (12.840, +9.267) (12.860, +9.260) (12.880, +9.274) (12.900, +9.276) (12.920, +9.275) (12.940, +9.257) (12.960, +9.275) (12.980, +9.287) (13.000, +9.286) (13.020, +9.277) (13.040, +9.270) (13.060, +9.275) (13.080, +9.267) (13.100, +9.267) (13.120, +9.263) (13.140, +9.262) (13.160, +9.258) (13.180, +9.265) (13.200, +9.263) (13.220, +9.270) (13.240, +9.263) (13.260, +9.273) (13.280, +9.273) (13.300, +9.270) (13.320, +9.265) (13.340, +9.259) (13.360, +9.259) (13.380, +9.252) (13.400, +9.258) (13.420, +9.261) (13.440, +9.261) (13.460, +9.269) (13.480, +9.264) };
                \draw [densely dotted, gray] (5, 9.25) -- (13.5, 9.25);

            \end{tikzpicture}
            \caption{Schematic visualization of the proposed architecture to illustrate the effect of dilated convolutions on the receptive field: the highlighted (solid orange) value in the fourth layer depends on exactly 8 values in the input layer.
            It therefore has an receptive field of size 8.
            The figure also shows how the length of the time input is
            successively reduced with each convolutional layer: the output of layer $i$ is $r_i-1$ time steps shorter than the original input, where $r_i$ denotes the receptive field of that layer.}
            \label{fig:model-architecture}
        \end{figure*}
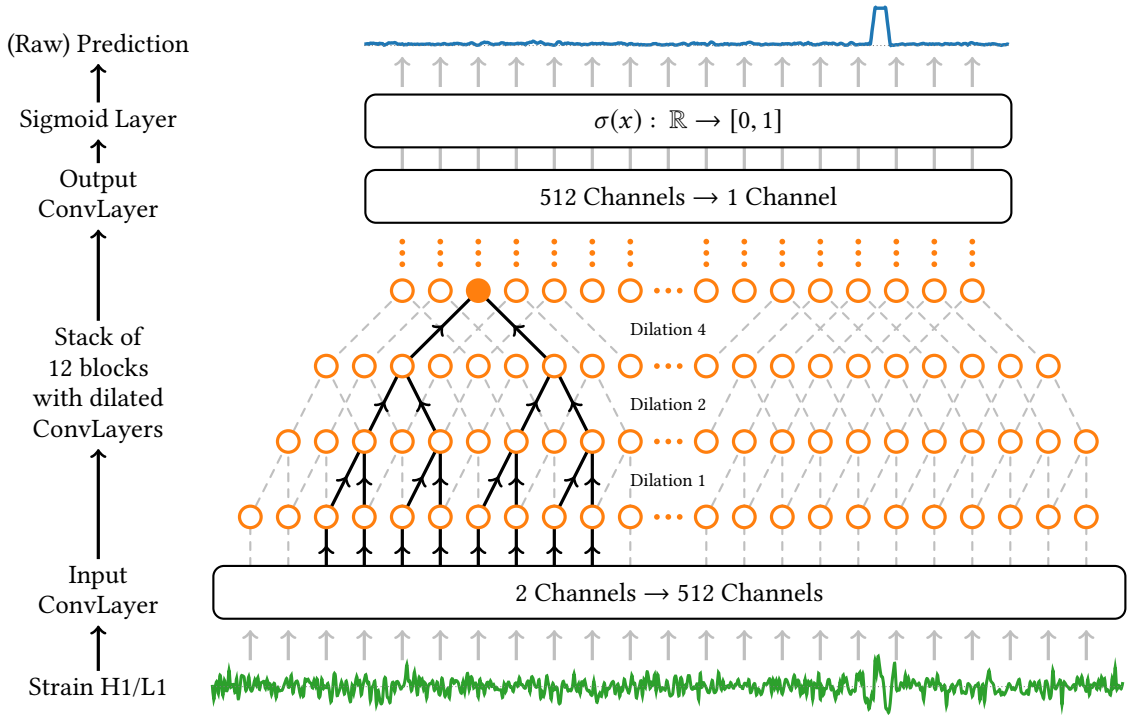

        In this section, we develop our specific neural network architecture (which aims to avoid some of the previously mentioned problems of CNNs) and document the training procedure.
        A high-level schematic drawing of the model architecture is depicted in \cref{fig:model-architecture}.

        \subsection{Model architecture}
        \label{subsec:model-architecture}

        In order to achieve a model that is agnostic to the length of the input time series, we choose a \emph{fully convolutional} architecture.
        This means there are no fully connected (or dense) layers.
        Instead, the neural network only learns convolutional filters (or kernels), which make no assumptions about the size of their input data.

        This has two major advantages. 
        First, if the size of the \emph{receptive field} of the network is $r$, we can directly evaluate our model on a time series of $n$ time steps for any $n > r$, resulting in an output time series of length $n - r + 1$.
        The receptive field of a network refers to the number of time steps on the input layer that affect a single time step on the output layer.
        Typically, an architecture should be chosen such that the receptive field is large enough to cover a substantial part of the signal.
        Second, it is more computationally efficient than a sliding window approach, which---due to the overlap of neighboring windows---performs redundant computations.
        A fully convolutional architecture avoids this overhead.

        Moreover, instead of evaluating the network for each detector separately, we stack the recordings from all observatories and treat them as different channels of a single, multidimensional input.
        This means that when the number of detectors changes, we only need to adjust the number of input channels of the first layer, while the rest of the architecture remains fixed.
        While retraining is required after such an extension, the computational complexity of our approach at test time is virtually constant in the number of detectors.

        In practice, we use a stack of \num{12} \emph{(convolutional) blocks}, each based on a dilated convolutional layer with \num{512} convolutional kernels of size \num{2}.
        Empirically, we found that increasing the number of channels used in the convolutional blocks generally improves the overall performance.
        However, memory limitations during training upper-bounded the number of channels to 512.
        Within each block, the convolutional layer itself is followed by a non-linear activation function, namely a rectified linear unit (ReLU).
        We did not use any regularization techniques such as dropout or batch normalization.

        The difference between the twelve convolutional blocks is the dilation of the kernels, which increases exponentially in powers of two (\ie $1,\, 2,\, 4,\, \ldots,\, 2048$) with the block number.
        This simple trick yields a relatively large receptive field of 2 seconds with a moderate depth of only 12 blocks while avoiding loss of resolution or coverage.
        This was considered sufficient to cover the relevant region around the
        coalescence for all signals of interest.
        Other modifications of the kernel, such as strides, were not used.

        The stack of convolutional blocks is preceded by an \emph{input} convolutional layer with kernel size of 1, which maps the input data from two channels (the strains from H1 and L1) to 512 channels.
        On the output side of the network, the last convolutional block is succeeded by an \emph{output} convolutional layer, which again has a kernel size of 1 and serves to reduce the number of channels from 512 back to 1.
        The now one-dimensional network output is then passed through a sigmoid layer \footnote{The sigmoid activation function $\sigma: \mathbb{R} \to (0,1)$ is defined as $\sigma(x) := 1 / (1 + e^{-x})$.}, which maps it into the interval $(0, 1)$.

        Our implementation (in Python~3.6.7) is based on the \textsc{PyTorch} deep learning framework (version~1.0.1) \cite{Paszke_2017}.
        All code that was used to obtain the results presented in this work is available online at \cite{Gebhard_2019b}.

        \subsection{Training procedure}
        \label{subsec:training-procedure}

        As usual for CNNs, before feeding an example time series $x$ as input during training, validation, and test time, we normalize it via $(x - \mu) / \sigma$, where $\mu$ and $\sigma$ are computed as the medians of the mean and standard deviation of each individual example in the training set.
        During training, we monitor the generalization performance by regularly evaluating the model on the validation set.
        For the actual training, we first use the \emph{Kaiming} initialization scheme as introduced in~\cite{He_2015} to assign initial random values to the network parameters (\ie the convolutional kernels).
        During training, the kernel entries are optimized using stochastic gradient descent using \emph{Adam}~\cite{Kingma_2014} with the \emph{AMSgrad} modification proposed in~\cite{Reddi_2018}.
        To this end, within every \emph{epoch} (\ie a full pass over all training data) the entire training dataset is randomly shuffled and divided into a fixed number of \emph{minibatches}.
        We use \emph{binary cross-entropy} (BCE) as the loss function.
        The \emph{batch loss} is calculated as the average of the BCE losses at every time step of every example in the minibatch and its corresponding label value.
        This batch loss is then automatically differentiated with respect to all kernels, and error back-propagation is used to update the kernel values.

        At the end of every epoch, the performance of the network with its current parameter values is evaluated both on the full training and validation data set.
        The current loss (as well as other metrics, see below) are logged and a \emph{checkpoint} of the model is created.
        This means that a copy of the model parameters is saved to disk such that the current training state can later be loaded again.
        We end training after a fixed number of epochs and retrieve the checkpoint corresponding to the lowest validation loss as the final \emph{trained model}.
        This is a form of \emph{validation-based early stopping}, which helps to avoid overfitting.

        By default, we choose an initial learning rate of $\eta_\text{init} = \num{3e-4}$.
        During training, the learning rate is reduced whenever the loss on the validation set has not decreased by more than a certain threshold over a given number of epochs (default value: 8).
        This behavior is controlled by \textsc{PyTorch}'s \hltt{ReduceLROnPlateau} method.

        In practice, we have trained our network for 64 epochs on the full training set using 5 NVIDIA Tesla V100 GPUs, each with \SI{32}{\giga\byte} of memory. 
        In total, training the model took approximately 30 hours on our hardware.
        This was deemed sufficient, as the network started to show signs of overfitting after approximately 30 epochs.
        As mentioned above, however, at test time (\ie for all evaluation experiments) we only used the model checkpoint with the lowest validation loss.

        \subsection{Postprocessing}
        \label{subsec:postprocessing}

        Finally, we apply two postprocessing steps to the raw network output: \emph{smoothing} and \emph{thresholding}.

        To \emph{smooth} the output time series, we apply a rolling average as a convolution with a rectangle function.
        The window size (\ie width) of this rectangle function can be tuned depending on the metric we want to optimize (see next section).
        Smoothing removes short spikes, which otherwise could be confused with the presence of signals.
        By default, we choose a window size of 256 time steps.

        In the subsequent \emph{thresholding} step, the smoothed output is mapped from $(0, 1)$ to $\{0,1\}$ depending on whether it exceeds a threshold $t$.
        This allows for stable and efficient peak-finding (see next section).
        Again, the choice of the threshold $t$ depends on the metric that one ultimately wants to optimize.
        By default, we used $t=0.5$.

        Both postprocessing steps are only applied at \emph{test time}, and we evaluate the effect of the parameter choices on the final performance in \cref{sec:experiments-and-results}.
        To compute the loss during training, we only use the raw, nonprocessed output of the network.

        \section{Performance Metrics}
        \label{sec:performance-metrics}

        \subsection{Design and creation of labels}
        \label{subsec:design-and-creation-of-labels}

        Let us now explain how we generate the \emph{labels}, that is, the desired network output for a given input.
        In our case, the labels are also time series:
        Ideally, the network should mark the exact locations of coalescences. 
        A natural way to represent this is a time series which is zero everywhere except at the event time where the signal in H1 reaches its maximum amplitude (where the label takes on a value of $1$).

        From a practical machine learning point of view, however, this is problematic:
        such sparse signals do not contribute sufficiently to the average loss to keep the network from simply \emph{always} predicting zero.
        To prevent this failure mode, instead of labeling a single time point, we choose a fixed-width interval centered around the time when the injected signal in the H1 channel reaches its maximum amplitude.
        By construction of our data generation pipeline, this position is fixed for all examples.
        (Recall that our fully convolutional network architecture is by design unable to overfit to specific locations within input examples.)
        Thus, labels need not be pre-generated or stored, but can be computed on the fly during training or testing.
        By default, the labels width (\ie the length of the symmetric interval around the event time in which the label time series takes on a value of $1$) is \SI{0.2}{\second}.

        \subsection{Evaluation metrics at test time}
        \label{subsec:evaluation-metrics-at-test-time}

        In \cref{sec:problem-setup-and-related-work} we discussed the drawbacks of the true positive rate and the false positive rate as performance measures for gravitational-wave searches in the example based binary classification setup.
        The fact that our data generation pipeline also generates individual examples is merely to make training convenient.
        Our model could equivalently be trained on a single time series (of sufficient length) containing any number of injections at arbitrary locations.
        This is possible because our architecture does not perform binary classification on an example level, but outputs yet another time series.
        As a consequence, different performance metrics are required.

        Our objective is to tag signals in the data by outputting a peak \emph{close} to the actual coalescence time.
        This intuition can be formalized to obtain interpretable performance metrics using the following evaluation procedure:

        \begin{enumerate}
            \item We identify all intervals of value 1 in the smoothed and thresholded network output.
            \item The interval centers are stored as \emph{candidate times}.
            \item For each candidate time $t_c$, we test for coincidence with the ground truth injection time $t_i$, that is, if \mbox{$|t_c - t_i| \leq \Delta t$}.
            By default, we use $\Delta t = \SI{0.05}{\second}$.
            Note that $\Delta t$ is another free, tuneable hyperparameter whose value will directly affect the performance metrics defined below.
            \item If a candidate time passes this coincidence check, we count it as a \emph{true positive} or \emph{detection} (see note below); otherwise, it is a \emph{false positive}.
            \item Per example, there can only be one true positive.
            If multiple candidate times pass the coincidence test for a single example, only one of them is counted as a detection, while the others are false positives.
        \end{enumerate}

        \begingroup
        \renewcommand{\quote}{\list{}{\rightmargin=0pt \topsep=0.6cm}\item\relax}
        \itshape
        \begin{quote}
            \textbf{Note:} We use the term \emph{detection} in this context to refer to an injected signal which was successfully recovered (in the sense of the procedure described above) by the network.
            This is, however, purely for ease of terminology. 
            A \enquote{detection} by the CNN cannot be compared to and must not be confused with the (statistically significant) detection of a gravitational wave as described in \cref{subsec:pycbc}.
            Similarly, the \emph{false positive rate} (see below) cannot directly be compared to a \emph{false alarm rate}.
        \end{quote}
        \endgroup

        We can now discuss the network performance on the test set in terms of the \emph{detection ratio} and the \emph{false positive ratio}.
        The detection ratio is simply the number of injected signals in the test set that the network could recover, divided by the total number of injected signals.
        We therefore also call it \emph{sensitivity}.
        The false positive ratio is the number of false positives divided by the number of all produced candidate times.
        It is thus an estimate of the error probability; the probability that any given candidate time does not coincide with an actual signal.

        Additionally, we can also define the \emph{false positive rate}:
        the total number of false positives divided by the combined duration of all samples in the test set.
        Its inverse is more intuitive: the \emph{inverse false positive rate} is the average time between two false positives.
        Naturally, this number should be as high as possible, meaning false positives should be as infrequent as possible.

        Again, note that our metrics do not rely on the existence of distinct examples, but could equally be evaluated on a single time series of arbitrary length containing multiple signals.
        To illustrate this key difference further, let us go back to our argument why the true positive rate and the false positive rate cannot be used to evaluate example based binary classification approaches in the sliding window mode of operation considering the output sequence $1 - 1 - 1 - 0 - 1 - 1 - 0$.
        First, previous approaches do not explain how to interpret such an undesirable situation.
        Moreover, their performance metrics are blind to these occurrences, because they are derived only from fixed-length examples, which all have an unambiguous binary label.
        Taking into account the continuous nature of the task, our metrics acknowledge this issue by counting at least one of the two positive intervals as a false positive if there was only one real signal within the corresponding time interval.

        \section{Experiments and Results}
        \label{sec:experiments-and-results}

        \subsection{Performance evaluation}
        \label{subsec:performance-evaluation}

        When evaluated on our full test set using the default parameters, our trained model is able to successfully recover approximately 89\% of all injections, while on average producing a false positive about once every 19.5 minutes.

\begin{figure}
            \centering
            \includegraphics[width=\linewidth]{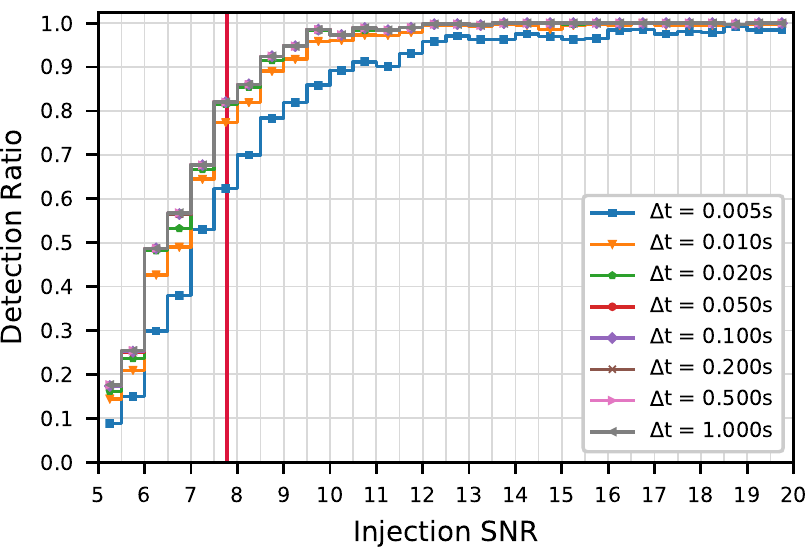}
            \caption{The detection ratio (DR) for positive examples binned by their network injection SNR (shown for different values of $\Delta t$).
            The DR increases steeply and plateaus at essentially 100\% for an SNR\,\raisebox{0.3mm}{$\scriptstyle\gtrsim$}\,$11$ (for $\Delta t\, \raisebox{0.3mm}{$\scriptstyle\geq$}\, \SI{0.01}{\second}$).
            The vertical red line indicates the network SNR threshold above which the \pycbc{} search pipeline considers events for further analysis.}
            \label{fig:main-results}
            \vspace{10.875mm}
            \includegraphics[width=\linewidth]{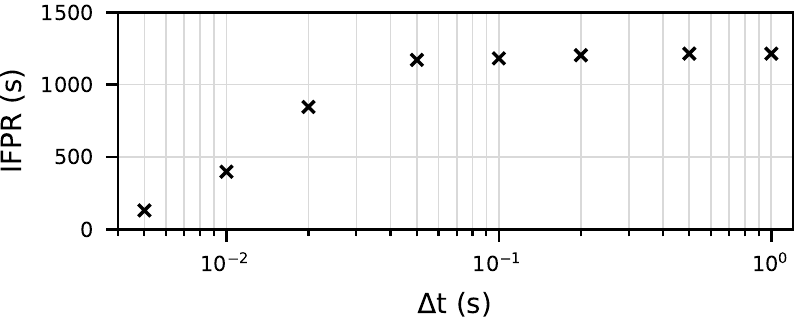}
            \caption{The inverse false positive rate (IFPR) as a function of the parameter~$\Delta t$ that controls how much a predicted event time~$t_c$ may deviate from the ground truth injection time~$t_i$ to still be counted as a detection (see step 3 in \cref{subsec:evaluation-metrics-at-test-time}).}
            \label{fig:ifpr-over-delta-t}
            \vspace{10.875mm}
            \includegraphics[width=\linewidth]{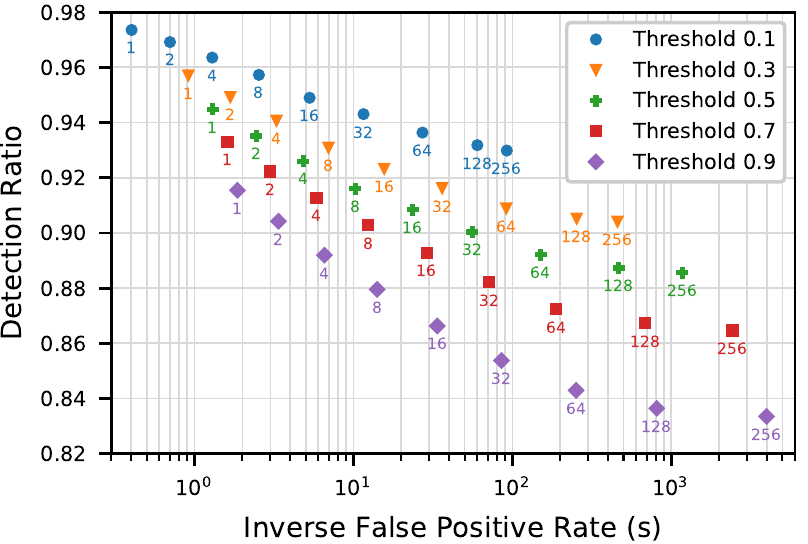}
            \caption{This figure shows the effect of the smoothing and thresholding parameters used during the postprocessing step on the detection ratio and the inverse false positive rate.
            Symbols encode the different threshold values, while the number next to the data points indicates the size of the smoothing window.
            The plot shows that these two parameters provide interpretable tuning knobs to choose an operating point.}
            \label{fig:postprocessing-effects}
            \vspace{-4.17671pt}
        \end{figure}

        For a more fine grained analysis, we then split the positive examples (\ie the ones that do contain an injection) in the test set into 30 bins based on their respective injection SNR.
        The bins are distributed equidistantly and cover the full injection SNR range of $(5.0, 5.5), (5.5, 6.0), \ldots , (19.5, 20)$.
        On average, every bin therefore contains $0.75 \cdot \num{16384} / 30 \approx 410$ examples.
        We then compute the detection ratio independently for each of these bins using different values of $\Delta t$ to investigate how the sensitivity of our method scales with the faintness of the signals as well as a function of $\Delta t$.
        The results in \cref{fig:main-results} show that the detection ratio increases steeply with the injection SNR and achieves essentially 100\% roughly at an SNR of 11 for $\Delta t\,\geq\,\SI{0.01}{\second}$.
        Furthermore, we find that the value of $\Delta t$ only has a very moderate influence on the performance of the model: for all values $\Delta t\,\geq\,\SI{0.05}{\second}$, the results are virtually indistinguishable.

        For comparison, the threshold for a coincident trigger to even be analyzed within the \pycbc{} search pipeline is $\sqrt{5.5^2 + 5.5^2} = 7.79$.
        At this injection SNR, our model (using $\Delta t = \SI{0.05}{\second}$) already recovers more than 80\% of all injected signals.
        Furthermore, the first ten real binary black hole mergers observed so far had network SNRs between \num{9.5} and \num{30.9}~\citep{Abbott_2018-12-16}, which is well within the region in which our model has a virtually perfect detection ratio.

        Additionally, we also compute the \emph{global} inverse false positive rate (\ie averaged over all examples) as a function of $\Delta t$.
        We show the results for this in \cref{fig:ifpr-over-delta-t}.
        For values $\Delta\,\geq\,\SI{0.05}{\second}$, the IFPR is virtually constant, which motivates our choice for the default value (\ie $\Delta t = \SI{0.05}{\second}$).

        \begin{figure*}
            \centering
            \subfloat[Results for GW150914.]{%
                \centering
                \includegraphics[width=0.995\linewidth]%
                    {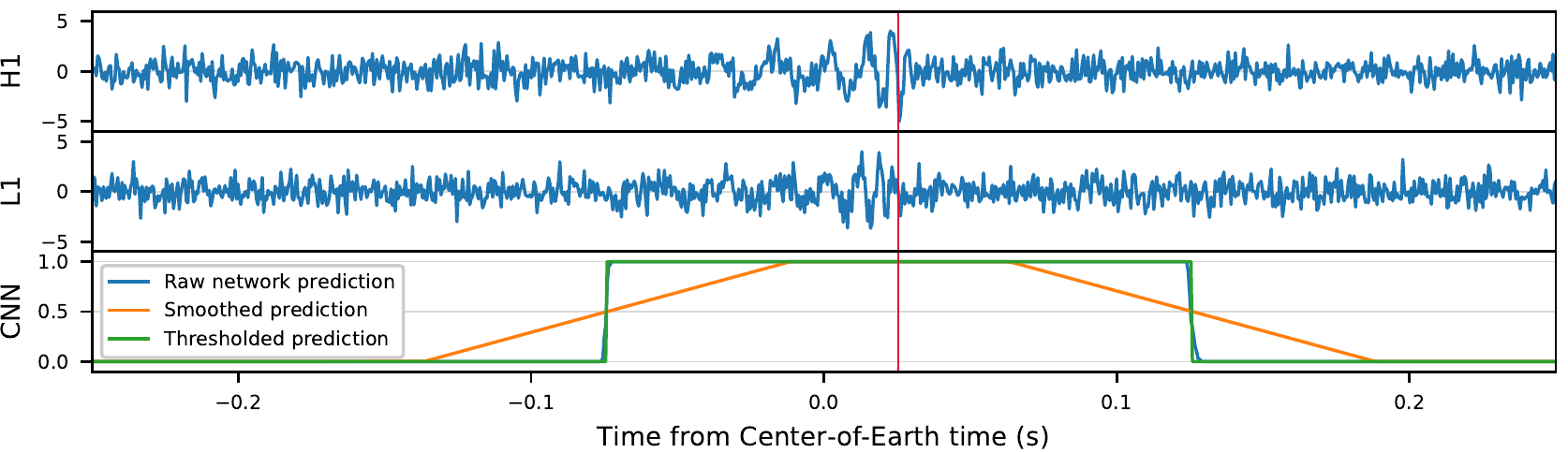}
            } %

            \subfloat[Results for GW151226.]{%
                \centering
                \includegraphics[width=0.995\linewidth]%
                    {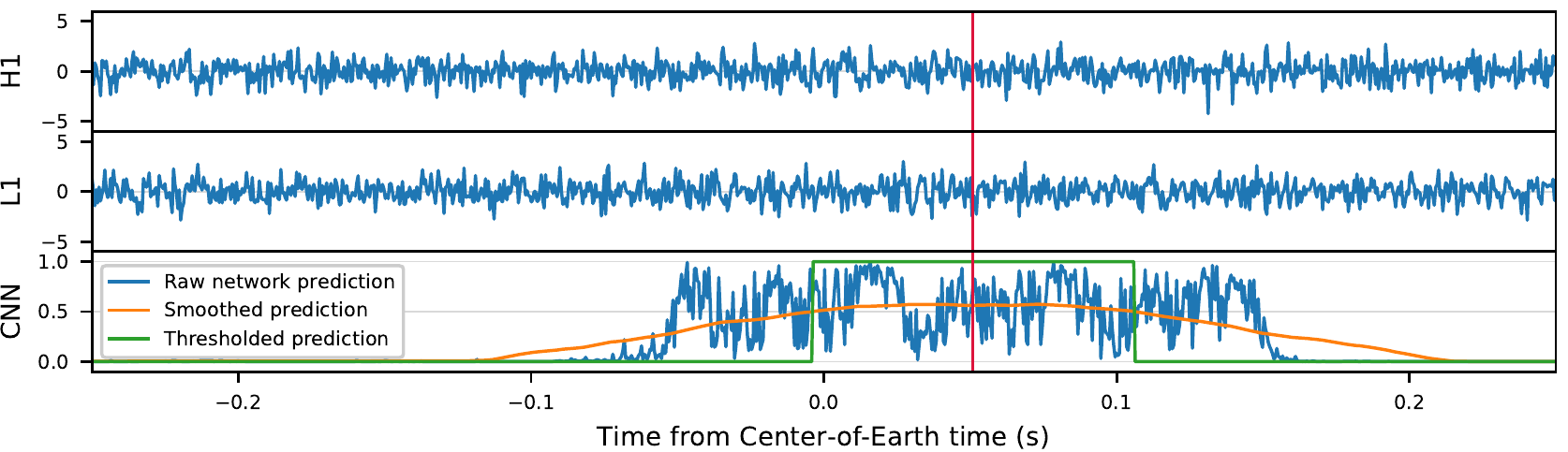}
            }
            \caption{Results for recovering the first two confirmed real events in \OR1, GW150914 and GW151226.
            The top two panels of each plot show the whitened, normalized strain for H1 and L1, centered around the time at which the peak of the gravitational-wave amplitude passed through the center of the Earth.
            The last panel shows the different postprocessing stages, namely, the raw, smoothed and thresholded network output (smoothing window size 256, threshold 0.5).
            The vertical red line indicates the predicted position of the event, calculated as the center of the interval of ones in the thresholded output.}
            \label{fig:real-event}
        \end{figure*}

        \subsection{Effects of postprocessing}
        \label{subsec:effects-of-postprocessing}

        Next, we systematically investigate the effect of both the smoothing and thresholding parameters.
        To this end, we postprocess the raw network output on the test set with different sizes of the smoothing window (1, 2, 4, 8, 16, 32, 64, 128, and 256) and different thresholds (0.1, 0.3, 0.5, 0.7, and 0.9) using our default value for $\Delta t$.
        In the parametric plot in \cref{fig:postprocessing-effects}, we show the detection ratio and the inverse false positive rate averaged over the entire test set for each combination of parameter settings (meaning up and right are better).
        While there is no single best option, this plot shows that our two parameters provide clearly interpretable tuning knobs to choose an operating point by trading off the sensitivity and the false positive rate.
        Depending on the application requirements one may use this plot to optimize detection ratio at fixed false positive rate or vice versa.

        \subsection{Recovering real gravitational-wave events}
        \label{subsec:recovering-real-gravitational-wave-events}

        In the next experiment, we evaluate our model's ability to generalize from synthetic training data to real events.
        The first two observations announced during LIGO's first observation run were GW150914 and GW151226 \cite{Abbott_2016-02-11,Abbott_2016-06-15}.
        These real signals were not included in the training data.
        At test time, we select an interval centered around the event times from the original recordings for both events, and apply the established whitening and band-passing procedure.
        Both samples are then cropped to \SI{16}{\second}, again centered around the event time.
        After normalizing and passing them through the network, we apply our usual postprocessing steps, using a window size of 256 time steps for the smoothing and thresholding the result at \num{0.5}.

        The results in \cref{fig:real-event} show that in both cases, the model was able to successfully recover the real GW signal at the correct position despite being slightly less accurate on the fainter event GW151226 (with a network SNR of 13) than the first observed event GW150914 (with a network SNR of 24) \cite{Abbott_2016-02-11,Abbott_2016-06-15}.
        The fainter example highlights the effect of postprocessing: 
        Instead of causing multiple false positives when thresholding the raw network output directly, the additional smoothing step yields a single connected interval (\ie a single predicted event time).

        Finally, we also apply our trained network to all other events in the GWTC-1 catalog~\cite{Abbott_2018-12-16}, which consists of 11 confirmed binary mergers from both the first and second observation run of LIGO. 
        Using the event data available from the GWOSC (which was preprocessed in the same way as before), we find that our network can indeed recover all known events, with the exception of GW170817.
        This is, however, not a surprise: While all other events are binary \emph{black hole} mergers, and we also trained our model using simulated BBH waveforms, GW170817 is the only confirmed binary \emph{neutron star} merger~\cite{Abbott_2017-10-16a}.

        Lastly, the fact that we are able to also successfully recover the events from \OR2 after using only recordings from \OR1 to train also indicates that the model is, to a certain extent, robust to changes in the detector characteristics.

        \subsection{A note of caution}
        \label{subsec:a-note-of-caution}

        \begin{figure*}
            \centering
            \subfloat[Examples that visually seem to resemble a gravitational-wave signal (\ie chirp-like increase in frequency and amplitude).]{%
                \centering
                \includegraphics[width=5.5cm]
                    {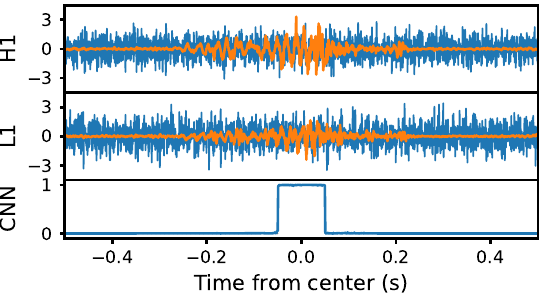}%
                \hspace{0.25cm}%
                \includegraphics[width=5.5cm]
                    {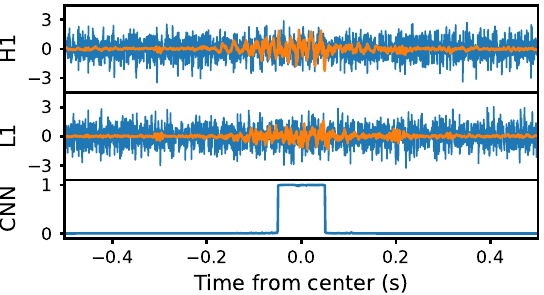}%
                \hspace{0.25cm}%
                \includegraphics[width=5.5cm]
                    {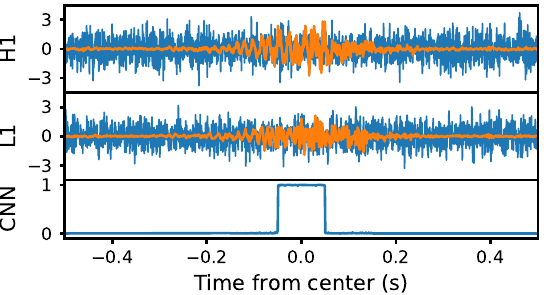}%
            } %

            \subfloat[Examples where no clear chirp-like pattern is visually discernible.]{%
                \centering
                \includegraphics[width=5.5cm]
                    {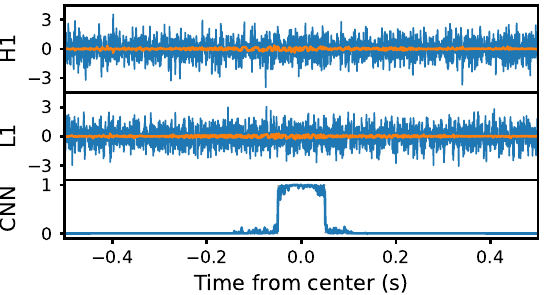}%
                \hspace{0.25cm}%
                \includegraphics[width=5.5cm]
                    {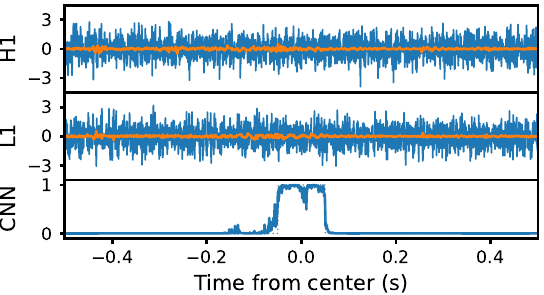}%
                \hspace{0.25cm}%
                \includegraphics[width=5.5cm]
                    {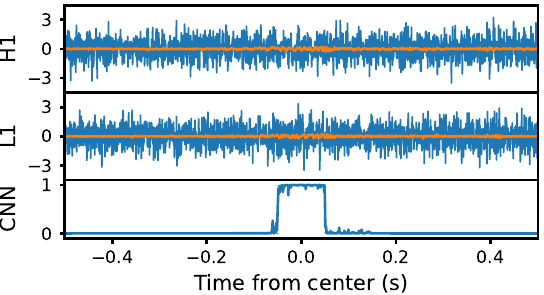}%
            } %

            \subfloat[Examples which satisfy unphysical constraints, yet still cause the network to predict the presence of a signal.
            In the first example, the input strain is constrained to only non-negative values.
            In the second example, the input strain is constrained to 0 in the \SI{0.25}{\second}-interval around predicted event time.
            In the last example, the entire example is constrained to have a minimal strain amplitude.]{%
                \centering
                \includegraphics[width=5.5cm]
                    {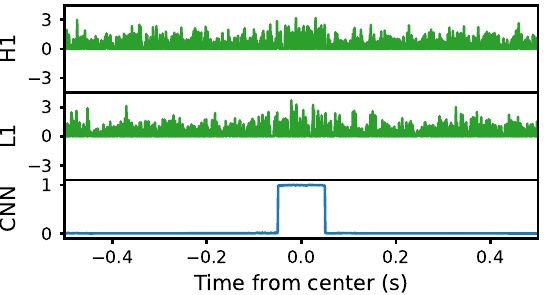}%
                \hspace{0.25cm}%
                \includegraphics[width=5.5cm]
                    {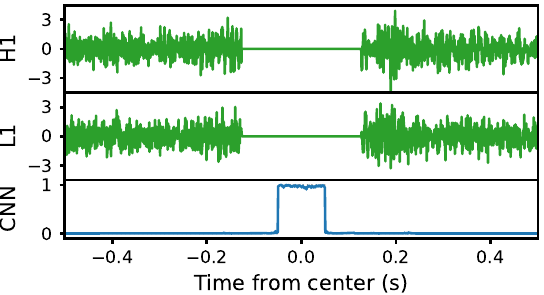}%
                \hspace{0.25cm}%
                \includegraphics[width=5.5cm]
                    {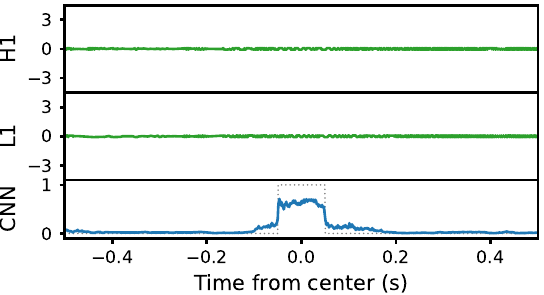}%
            }
            \caption{This figures shows different example results where we---using a fixed pretrained model---optimized the network inputs (starting from noise-only examples) in order to produce a given desired output.
            The top and middle panel show the strain for the two detectors, H1 and L1.
            The original inputs (\ie the pure background noise) are shown in blue, and the \emph{difference} between the original and the optimized input is shown in orange. 
            This is the component that is added to the noise in order to make the network predict the presence of a \enquote{signal}.
            Ideally, we would therefore expect the orange component to look like a gravitational-wave waveform.
            For the examples in subfigure c), only the the effective (\ie optimized and constrained) inputs to the network are shown (in green).
            The bottom panel of every figure shows the desired output (\ie the optimization target) in dotted gray, and the raw network prediction in blue (\ie without any postprocessing).}
            \label{fig:adversarial}
        \end{figure*}

        In a final experiment, we once more want to emphasize our call for caution when interpreting CNNs in the context of gravitational-wave searches.
        To address the question \enquote{What has the model actually learned?,} we use techniques inspired from \emph{activation maximization} or \emph{feature visualization} (see, \eg \cite{Zeiler_2014, Olah_2017}), as well as \emph{adversarial examples} or \emph{adversarial attacks} (see, \eg \cite{Szegedy_2013}), which are currently active areas of research within the machine learning community.
        Specifically, we perform the following test in which we make use of the differentiability of our model to find examples of inputs which cause the network to produce a given target output:

        \begin{enumerate}
            \item We randomly select a noise-only example (\ie an example that does not contain an injection) from our testing set and crop it from the end to a length of \SI{3}{\second}.
            This is our initial network input.
            \item Next, we generate a target label, which is about \SI{1}{\second} long (\SI{3}{\second} minus the receptive field of the model) and zero everywhere except for the interval from \SI{0.45}{\second} to \SI{0.55}{\second}, where it takes on a value of 1.
            \item If applicable, we enforce additional constraints on the inputs.
            For example, we pass the input through a $\max(x, 0)$-function to create the physically nonsensical scenario of a strain that is strictly non-negative (see first example in \cref{fig:adversarial}~c).
            \item We pass the constrained network input through the trained model from the previous experiments.
            We then compute a weighted sum of a binary cross-entropy and a mean squared error loss between the network prediction and the target.
            The exact weighting depends on the optimization target.
            \item Unlike when training a neural network, this loss is then not back-propagated to the weights of the network, which stay fixed during this experiment. 
            Instead, the loss is back-propagated to the \emph{input}, which is updated in order to minimize the loss.
            \item We repeat this procedure (starting with enforcing possible constraints on the inputs) for 256 iterations, again using \emph{Adam} as the optimizer, with an initial learning rate of $\eta = 0.3$.
            \textsc{PyTorch}'s default cosine annealing scheduler is used to gradually decrease the learning rate every epoch.
            \item Finally, we compute the difference between the original network input and the optimized input.
            This can be interpreted as the hypothetical \enquote{signal}, which---when added into the pure noise example---makes our network produce the target output.
        \end{enumerate}

        We repeat this procedure for different initial inputs and manually inspect the results in form of the hypothetical \enquote{signals} to check if they match our expectation:
        If the network had truly learned to respond only to gravitational waves, we would expect these hypothetical signals to closely resemble gravitational-wave signals.

        However, while some of the inputs that have undergone the described optimization procedure do exhibit a chirplike structure (\ie oscillations increasing in both amplitude and frequency), we find that this is not always the case; see panel (a) and (b) of \cref{fig:adversarial}.
        Worse yet, we can also achieve the desired output even when imposing non-physical constraints on the inputs.
        We investigate three types of such constraints:
        First, we allow only non-negative strain values.
        Second, we enforce the strain to be zero in a \SI{0.25}{\second}-interval covering the interval in which the target output is one.
        Third, we clip the network input values to a small interval around zero to minimize the overall amplitude.
        In all three cases, we can still find examples that obey the constraints and, when passed through the network, yield the desired target output.
        Examples for this are shown in panel (c) of \cref{fig:adversarial}.

        Since we crafted these examples in a supervised fashion, one may argue that the cases in panel (c) are \emph{unrealistically} out of distribution, that is, they would never occur in real detector recordings and therefore do not lead to complications in practice.
        However, in particular the unconstrained examples in panel (b) of \cref{fig:adversarial} are unsettling, because they illustrate just how easily the network can be fooled even by small changes in the inputs.
        These results suggest a detailed quantification of how contrived these hypothetical signals really are (measured by how likely they are to occur accidentally in future detector recordings) to assess whether one must account for them in the false positive rate.
        Without such an analysis the worry of overconfident positive CNN output on pure noise or faint non-Gaussian transients remains.

        \section{Discussion and Conclusion}
        \label{sec:discussion-and-conclusion}

        In this work we provide an interdisciplinary, in-depth analysis of the potential of deep convolutional neural networks (CNNs) as part of the effort around searching for gravitational waves from binary coalescences in strain data.
        First, we critically scrutinize both the methods as well as the contributions of existing works on this topic by carefully analyzing how standard machine learning approaches and metrics map to the specific task at hand.
        This analysis yields two major conclusions:
        \begin{enumerate*}
            \item CNNs alone cannot be used to claim statistically significant gravitational-wave detections.
            \item Fast inference times, favorable computational scaling in the number of detectors, and a compact internal representation of a large number of waveforms presented during training still make CNNs a useful and promising tool to produce real-time triggers for detailed analysis and follow up searches.
        \end{enumerate*}

        As part of these key conceptual insights, we hope to foster further interdisciplinary research on this topic by highlighting important subtleties of GW searches to machine learning experts and exposing some potential pitfalls and surprising properties of CNNs to physicists.

        Building on these insights, we have designed a flexible data generation pipeline which we make publicly available as an open source package.
        We use a novel network architecture which is more tailored to the physical task at hand than a binary classification-based approach and also overcomes some subtle pitfalls, such as the danger of overfitting to some particular properties of the training data.
        We evaluate this approach on real LIGO recordings and demonstrate the potential of such a system as a trigger generator by achieving a detection ratio of 86\% with a false positive on average once every 40 minutes.
        Two tuneable postprocessing parameters allow us to intuitively trade off between the detection ratio and the false positive rate without having to retrain the model.

        Finally, as part of our effort for cross-disciplinary understanding, we showcase a selection of \enquote{failure modes} of our model which are typical for deep convolutional neural networks.
        We contrive inputs which the network believes to contain gravitational-wave signals with high confidence, even though they are structurally very different from real detector signals for compact binary coalescences.
        While some of these inputs are physically unrealistic and thus unlikely to be observed in practice, others appear quite plausible (\eg tiny modifications of pure noise examples).
        Because the detector noise properties change on an hourly timescale, the rate of false triggers due to such failures may be hard to predict even for a well-tuned CNN.
        We leave the required quantitative analysis of how such incidences may affect the performance on real-world recordings under changing detector characteristics for future research, and conclude this work with a note of caution: CNNs are a promising tool for gravitational-wave data analysis; however, their exact interpretation requires great care and attention.

    \appendix

        \begin{acknowledgments}
            We would like to thank Thomas Dent, Alexander Nitz, and Giambattista Parascandolo for suggestions and feedback on the manuscript.
            In addition, we want to thank the anonymous referee for carefully reading this manuscript and providing valuable feedback.
            T.\,D.\,G., N.\,K., I.\,H. and B.\,S. acknowledge the support of the Max-Planck-Gesell\-schaft.
            T.\,D.\,G. acknowledges partial funding from the Max Planck ETH Center for Learning Systems.
            This research has made use of data, software and/or web tools obtained from the
            Gravitational Wave Open Science Center (\url{https://www.gw-openscience.org}), a service of LIGO Laboratory, the LIGO Scientific Collaboration and the Virgo Collaboration.
            LIGO is funded by the U.S. National Science Foundation. 
            Virgo is funded by the French Centre National de Recherche Scientifique (CNRS), the Italian Istituto Nazionale della Fisica Nucleare (INFN) and the Dutch Nikhef, with contributions by Polish and Hungarian institutes.
        \end{acknowledgments}

        \section{Data Generation Parameters}
        \label{sec:data-generation-parameters}
        
        The following list explains the different parameters and the distributions from which their values are randomly sampled before being passed as inputs to the \hltt{SEOBNRv4} waveform model in order to simulate synthetic gravitational-wave signals.
        Because the true astrophysical distributions for compact binary coalescences are unknown, we choose the following generic values:
        
        \begin{itemize}[itemsep=0.5ex]
            \item \hltt{mass1} and \hltt{mass2}: The masses of the two merging black holes, chosen independently and uniformly at random between 10 and 80 solar masses.
            \item \hltt{spin1z} and \hltt{spin2z}: The $z$-component of the spin of the merging black holes, chosen independently and uniformly at random between 0 and 0.998 (to improve the numerical stability). 
            \item \hltt{ra} and \hltt{dec}: The right ascension of declination defining the position of the source in the sky.
                Both values are sampled together from a uniform distribution over the sky.
            \item \hltt{polarization}: The polarization angle is one of the three Euler angles relating the radiation frame, which is the reference frame in which the gravitational wave propagates in the $z$-direction, to the reference frame of the detector.
                It is sampled uniformly at random from the interval $[0, 2\pi]$.
            \item \hltt{coa_phase} and \hltt{inclination}: To understand the significance of the \emph{coalescence phase} and the \emph{inclination}, one needs to introduce a third reference frame beside the detector and radiation frame, namely, the reference frame of the source itself. 
                In the case of a binary coalescence, this source reference frame is chosen such that its $z$-axis is perpendicular to the plane in which the two black holes orbit each other. 
                Then, the \hltt{coa_phase} and the \hltt{inclination} are the two angles that specify the location in the sky of the detector as seen from this source frame.
                Their values are sampled jointly from a uniform distribution over a sphere.
            \item \hltt{injection_snr}: For evaluation purposes, it is useful to generate samples with a pre-defined signal-to-noise ratio.
                This can be achieved by re-scaling the waveform, which is physically equivalent to moving the source closer or further from the detector.
                The \hltt{injection_snr} is the desired network SNR for the example, which is sampled uniformly from $[5, 20]$.
                It is not directly passed to the simulation routine, but only used later when adding the simulated signal into the background noise.
        \end{itemize}

        \bibliography{magic-bullet.bib}

\begin{thebibliography}{76}%
\makeatletter
\providecommand \@ifxundefined [1]{%
 \@ifx{#1\undefined}
}%
\providecommand \@ifnum [1]{%
 \ifnum #1\expandafter \@firstoftwo
 \else \expandafter \@secondoftwo
 \fi
}%
\providecommand \@ifx [1]{%
 \ifx #1\expandafter \@firstoftwo
 \else \expandafter \@secondoftwo
 \fi
}%
\providecommand \natexlab [1]{#1}%
\providecommand \enquote  [1]{``#1''}%
\providecommand \bibnamefont  [1]{#1}%
\providecommand \bibfnamefont [1]{#1}%
\providecommand \citenamefont [1]{#1}%
\providecommand \href@noop [0]{\@secondoftwo}%
\providecommand \href [0]{\begingroup \@sanitize@url \@href}%
\providecommand \@href[1]{\@@startlink{#1}\@@href}%
\providecommand \@@href[1]{\endgroup#1\@@endlink}%
\providecommand \@sanitize@url [0]{\catcode `\\12\catcode `\$12\catcode
  `\&12\catcode `\#12\catcode `\^12\catcode `\_12\catcode `\%12\relax}%
\providecommand \@@startlink[1]{}%
\providecommand \@@endlink[0]{}%
\providecommand \url  [0]{\begingroup\@sanitize@url \@url }%
\providecommand \@url [1]{\endgroup\@href {#1}{\urlprefix }}%
\providecommand \urlprefix  [0]{URL }%
\providecommand \Eprint [0]{\href }%
\providecommand \doibase [0]{https://doi.org/}%
\providecommand \selectlanguage [0]{\@gobble}%
\providecommand \bibinfo  [0]{\@secondoftwo}%
\providecommand \bibfield  [0]{\@secondoftwo}%
\providecommand \translation [1]{[#1]}%
\providecommand \BibitemOpen [0]{}%
\providecommand \bibitemStop [0]{}%
\providecommand \bibitemNoStop [0]{.\EOS\space}%
\providecommand \EOS [0]{\spacefactor3000\relax}%
\providecommand \BibitemShut  [1]{\csname bibitem#1\endcsname}%
\let\auto@bib@innerbib\@empty
%</preamble>
\bibitem [{\citenamefont {Allen}\ \emph {et~al.}(2012)\citenamefont {Allen}
  \emph {et~al.}}]{Allen_2012}%
  \BibitemOpen
  \bibfield  {author} {\bibinfo {author} {\bibfnamefont {B.}~\bibnamefont
  {Allen}} \emph {et~al.},\ }\bibfield  {title} {\bibinfo {title} {{FINDCHIRP:
  An Algorithm for Detection of Gravitational Waves from Inspiraling Compact
  Binaries}},\ }\href {https://doi.org/10.1103/PhysRevD.85.122006} {\bibfield
  {journal} {\bibinfo  {journal} {Physical Review D}\ }\textbf {\bibinfo
  {volume} {85}} (\bibinfo {year} {2012})}\BibitemShut {NoStop}%
\bibitem [{\citenamefont {Babak}\ \emph {et~al.}(2013)\citenamefont {Babak}
  \emph {et~al.}}]{Babak_2013}%
  \BibitemOpen
  \bibfield  {author} {\bibinfo {author} {\bibfnamefont {S.}~\bibnamefont
  {Babak}} \emph {et~al.},\ }\bibfield  {title} {\bibinfo {title} {Searching
  for gravitational waves from binary coalescence},\ }\href
  {https://doi.org/10.1103/physrevd.87.024033} {\bibfield  {journal} {\bibinfo
  {journal} {Physical Review D}\ }\textbf {\bibinfo {volume} {87}} (\bibinfo
  {year} {2013})}\BibitemShut {NoStop}%
\bibitem [{\citenamefont {Usman}\ \emph {et~al.}(2016)\citenamefont {Usman}
  \emph {et~al.}}]{Usman_2016}%
  \BibitemOpen
  \bibfield  {author} {\bibinfo {author} {\bibfnamefont {S.~A.}\ \bibnamefont
  {Usman}} \emph {et~al.},\ }\bibfield  {title} {\bibinfo {title} {{The PyCBC
  search for gravitational waves from compact binary coalescence}},\ }\href
  {https://doi.org/10.1088/0264-9381/33/21/215004} {\bibfield  {journal}
  {\bibinfo  {journal} {Classical and Quantum Gravity}\ }\textbf {\bibinfo
  {volume} {33}} (\bibinfo {year} {2016})}\BibitemShut {NoStop}%
\bibitem [{\citenamefont {Messick}\ \emph {et~al.}(2017)\citenamefont {Messick}
  \emph {et~al.}}]{Messick_2017}%
  \BibitemOpen
  \bibfield  {author} {\bibinfo {author} {\bibfnamefont {C.}~\bibnamefont
  {Messick}} \emph {et~al.},\ }\bibfield  {title} {\bibinfo {title} {Analysis
  framework for the prompt discovery of compact binary mergers in
  gravitational-wave data},\ }\href
  {https://doi.org/10.1103/physrevd.95.042001} {\bibfield  {journal} {\bibinfo
  {journal} {Physical Review D}\ }\textbf {\bibinfo {volume} {95}} (\bibinfo
  {year} {2017})}\BibitemShut {NoStop}%
\bibitem [{\citenamefont {Aasi}\ \emph {et~al.}(2015)\citenamefont {Aasi} \emph
  {et~al.}}]{Aasi_2015}%
  \BibitemOpen
  \bibfield  {author} {\bibinfo {author} {\bibfnamefont {J.}~\bibnamefont
  {Aasi}} \emph {et~al.},\ }\bibfield  {title} {\bibinfo {title} {Advanced
  {LIGO}},\ }\href {https://doi.org/10.1088/0264-9381/32/7/074001} {\bibfield
  {journal} {\bibinfo  {journal} {Classical and Quantum Gravity}\ }\textbf
  {\bibinfo {volume} {32}} (\bibinfo {year} {2015})}\BibitemShut {NoStop}%
\bibitem [{\citenamefont {Acernese}\ \emph {et~al.}(2014)\citenamefont
  {Acernese} \emph {et~al.}}]{Acernese_2014}%
  \BibitemOpen
  \bibfield  {author} {\bibinfo {author} {\bibfnamefont {F.}~\bibnamefont
  {Acernese}} \emph {et~al.},\ }\bibfield  {title} {\bibinfo {title} {{Advanced
  Virgo: a second-generation interferometric gravitational wave detector}},\
  }\href {https://doi.org/10.1088/0264-9381/32/2/024001} {\bibfield  {journal}
  {\bibinfo  {journal} {Classical and Quantum Gravity}\ }\textbf {\bibinfo
  {volume} {32}} (\bibinfo {year} {2014})}\BibitemShut {NoStop}%
\bibitem [{\citenamefont {Abbott}\ \emph
  {et~al.}(2016{\natexlab{a}})\citenamefont {Abbott} \emph
  {et~al.}}]{Abbott_2016-02-11}%
  \BibitemOpen
  \bibfield  {author} {\bibinfo {author} {\bibfnamefont {B.~P.}\ \bibnamefont
  {Abbott}} \emph {et~al.},\ }\bibfield  {title} {\bibinfo {title}
  {{Observation of Gravitational Waves from a Binary Black Hole Merger}},\
  }\href {https://doi.org/10.1103/physrevlett.116.061102} {\bibfield  {journal}
  {\bibinfo  {journal} {Physical Review Letters}\ }\textbf {\bibinfo {volume}
  {116}} (\bibinfo {year} {2016}{\natexlab{a}})}\BibitemShut {NoStop}%
\bibitem [{\citenamefont {Abbott}\ \emph
  {et~al.}(2016{\natexlab{b}})\citenamefont {Abbott} \emph
  {et~al.}}]{Abbott_2016-06-15}%
  \BibitemOpen
  \bibfield  {author} {\bibinfo {author} {\bibfnamefont {B.~P.}\ \bibnamefont
  {Abbott}} \emph {et~al.},\ }\bibfield  {title} {\bibinfo {title} {{GW151226:
  Observation of Gravitational Waves from a 22-Solar-Mass Binary Black Hole
  Coalescence}},\ }\href {https://doi.org/10.1103/physrevlett.116.241103}
  {\bibfield  {journal} {\bibinfo  {journal} {Physical Review Letters}\
  }\textbf {\bibinfo {volume} {116}} (\bibinfo {year}
  {2016}{\natexlab{b}})}\BibitemShut {NoStop}%
\bibitem [{\citenamefont {Abbott}\ \emph
  {et~al.}(2017{\natexlab{a}})\citenamefont {Abbott} \emph
  {et~al.}}]{Abbott_2017-06-01}%
  \BibitemOpen
  \bibfield  {author} {\bibinfo {author} {\bibfnamefont {B.~P.}\ \bibnamefont
  {Abbott}} \emph {et~al.},\ }\bibfield  {title} {\bibinfo {title} {{GW170104:
  Observation of a 50-Solar-Mass Binary Black Hole Coalescence at Redshift
  0.2}},\ }\href {https://doi.org/10.1103/physrevlett.118.221101} {\bibfield
  {journal} {\bibinfo  {journal} {Physical Review Letters}\ }\textbf {\bibinfo
  {volume} {118}} (\bibinfo {year} {2017}{\natexlab{a}})}\BibitemShut {NoStop}%
\bibitem [{\citenamefont {Abbott}\ \emph
  {et~al.}(2017{\natexlab{b}})\citenamefont {Abbott} \emph
  {et~al.}}]{Abbott_2017-10-06}%
  \BibitemOpen
  \bibfield  {author} {\bibinfo {author} {\bibfnamefont {B.~P.}\ \bibnamefont
  {Abbott}} \emph {et~al.},\ }\bibfield  {title} {\bibinfo {title} {{GW170814:
  A Three-Detector Observation of Gravitational Waves from a Binary Black Hole
  Coalescence}},\ }\href {https://doi.org/10.1103/physrevlett.119.141101}
  {\bibfield  {journal} {\bibinfo  {journal} {Physical Review Letters}\
  }\textbf {\bibinfo {volume} {119}} (\bibinfo {year}
  {2017}{\natexlab{b}})}\BibitemShut {NoStop}%
\bibitem [{\citenamefont {Abbott}\ \emph
  {et~al.}(2017{\natexlab{c}})\citenamefont {Abbott} \emph
  {et~al.}}]{Abbott_2017-12-18}%
  \BibitemOpen
  \bibfield  {author} {\bibinfo {author} {\bibfnamefont {B.~P.}\ \bibnamefont
  {Abbott}} \emph {et~al.},\ }\bibfield  {title} {\bibinfo {title} {{GW170608:
  Observation of a 19 Solar-mass Binary Black Hole Coalescence}},\ }\href
  {https://doi.org/10.3847/2041-8213/aa9f0c} {\bibfield  {journal} {\bibinfo
  {journal} {The Astrophysical Journal}\ }\textbf {\bibinfo {volume} {851}}
  (\bibinfo {year} {2017}{\natexlab{c}})}\BibitemShut {NoStop}%
\bibitem [{\citenamefont {Abbott}\ \emph
  {et~al.}(2018{\natexlab{a}})\citenamefont {Abbott} \emph
  {et~al.}}]{Abbott_2018-12-16}%
  \BibitemOpen
  \bibfield  {author} {\bibinfo {author} {\bibfnamefont {B.~P.}\ \bibnamefont
  {Abbott}} \emph {et~al.},\ }\href@noop {} {\bibinfo {title} {{GWTC-1: A
  Gravitational-Wave Transient Catalog of Compact Binary Mergers Observed by
  LIGO and Virgo during the First and Second Observing Runs}}} (\bibinfo {year}
  {2018}{\natexlab{a}}),\ \Eprint {https://arxiv.org/abs/1811.12907}
  {arXiv:1811.12907} \BibitemShut {NoStop}%
\bibitem [{\citenamefont {Abbott}\ \emph
  {et~al.}(2018{\natexlab{b}})\citenamefont {Abbott} \emph
  {et~al.}}]{Abbott_2019-01-03}%
  \BibitemOpen
  \bibfield  {author} {\bibinfo {author} {\bibfnamefont {B.~P.}\ \bibnamefont
  {Abbott}} \emph {et~al.},\ }\href@noop {} {\bibinfo {title} {{Binary Black
  Hole Population Properties Inferred from the First and Second Observing Runs
  of Advanced LIGO and Advanced Virgo}}} (\bibinfo {year}
  {2018}{\natexlab{b}}),\ \Eprint {https://arxiv.org/abs/1811.12940v3}
  {arXiv:1811.12940v3} \BibitemShut {NoStop}%
\bibitem [{\citenamefont {Abbott}\ \emph
  {et~al.}(2017{\natexlab{d}})\citenamefont {Abbott} \emph
  {et~al.}}]{Abbott_2017-10-16a}%
  \BibitemOpen
  \bibfield  {author} {\bibinfo {author} {\bibfnamefont {B.~P.}\ \bibnamefont
  {Abbott}} \emph {et~al.},\ }\bibfield  {title} {\bibinfo {title} {{GW170817:
  Observation of Gravitational Waves from a Binary Neutron Star Inspiral}},\
  }\href {https://doi.org/10.1103/physrevlett.119.161101} {\bibfield  {journal}
  {\bibinfo  {journal} {Physical Review Letters}\ }\textbf {\bibinfo {volume}
  {119}} (\bibinfo {year} {2017}{\natexlab{d}})}\BibitemShut {NoStop}%
\bibitem [{\citenamefont {Abbott}\ \emph
  {et~al.}(2017{\natexlab{e}})\citenamefont {Abbott} \emph
  {et~al.}}]{Abbott_2017-10-16b}%
  \BibitemOpen
  \bibfield  {author} {\bibinfo {author} {\bibfnamefont {B.~P.}\ \bibnamefont
  {Abbott}} \emph {et~al.},\ }\bibfield  {title} {\bibinfo {title}
  {{Multi-messenger Observations of a Binary Neutron Star Merger}},\ }\href
  {https://doi.org/10.3847/2041-8213/aa91c9} {\bibfield  {journal} {\bibinfo
  {journal} {The Astrophysical Journal}\ }\textbf {\bibinfo {volume} {848}}
  (\bibinfo {year} {2017}{\natexlab{e}})}\BibitemShut {NoStop}%
\bibitem [{\citenamefont {Abbott}\ \emph
  {et~al.}(2017{\natexlab{f}})\citenamefont {Abbott} \emph
  {et~al.}}]{Abbott_2017-10-16}%
  \BibitemOpen
  \bibfield  {author} {\bibinfo {author} {\bibfnamefont {B.~P.}\ \bibnamefont
  {Abbott}} \emph {et~al.},\ }\bibfield  {title} {\bibinfo {title} {{A
  gravitational-wave standard siren measurement of the Hubble constant}},\
  }\href {https://doi.org/10.1038/nature24471} {\bibfield  {journal} {\bibinfo
  {journal} {Nature}\ }\textbf {\bibinfo {volume} {551}} (\bibinfo {year}
  {2017}{\natexlab{f}})}\BibitemShut {NoStop}%
\bibitem [{\citenamefont {Abbott}\ \emph
  {et~al.}(2018{\natexlab{c}})\citenamefont {Abbott} \emph
  {et~al.}}]{Abbott_2018-10-15}%
  \BibitemOpen
  \bibfield  {author} {\bibinfo {author} {\bibfnamefont {B.~P.}\ \bibnamefont
  {Abbott}} \emph {et~al.},\ }\bibfield  {title} {\bibinfo {title}
  {{{GW}170817: Measurements of Neutron Star Radii and Equation of State}},\
  }\href {https://doi.org/10.1103/physrevlett.121.161101} {\bibfield  {journal}
  {\bibinfo  {journal} {Physical Review Letters}\ }\textbf {\bibinfo {volume}
  {121}} (\bibinfo {year} {2018}{\natexlab{c}})}\BibitemShut {NoStop}%
\bibitem [{\citenamefont {Aso}\ \emph {et~al.}(2013)\citenamefont {Aso} \emph
  {et~al.}}]{Aso_2013}%
  \BibitemOpen
  \bibfield  {author} {\bibinfo {author} {\bibfnamefont {Y.}~\bibnamefont
  {Aso}} \emph {et~al.},\ }\bibfield  {title} {\bibinfo {title} {Interferometer
  design of the {KAGRA} gravitational wave detector},\ }\href
  {https://doi.org/10.1103/physrevd.88.043007} {\bibfield  {journal} {\bibinfo
  {journal} {Physical Review D}\ }\textbf {\bibinfo {volume} {88}} (\bibinfo
  {year} {2013})}\BibitemShut {NoStop}%
\bibitem [{\citenamefont {Abbott}\ \emph
  {et~al.}(2018{\natexlab{d}})\citenamefont {Abbott} \emph
  {et~al.}}]{Abbott_2018-04-26}%
  \BibitemOpen
  \bibfield  {author} {\bibinfo {author} {\bibfnamefont {B.~P.}\ \bibnamefont
  {Abbott}} \emph {et~al.},\ }\bibfield  {title} {\bibinfo {title} {{Prospects
  for observing and localizing gravitational-wave transients with Advanced
  LIGO, Advanced Virgo and KAGRA}},\ }\href
  {https://doi.org/10.1007/s41114-018-0012-9} {\bibfield  {journal} {\bibinfo
  {journal} {Living Reviews in Relativity}\ }\textbf {\bibinfo {volume} {21}}
  (\bibinfo {year} {2018}{\natexlab{d}})}\BibitemShut {NoStop}%
\bibitem [{\citenamefont {Veitch}\ \emph {et~al.}(2015)\citenamefont {Veitch}
  \emph {et~al.}}]{Veitch_2015}%
  \BibitemOpen
  \bibfield  {author} {\bibinfo {author} {\bibfnamefont {J.}~\bibnamefont
  {Veitch}} \emph {et~al.},\ }\bibfield  {title} {\bibinfo {title} {Parameter
  estimation for compact binaries with ground-based gravitational-wave
  observations using the {LALInference} software library},\ }\href
  {https://doi.org/10.1103/physrevd.91.042003} {\bibfield  {journal} {\bibinfo
  {journal} {Physical Review D}\ }\textbf {\bibinfo {volume} {91}} (\bibinfo
  {year} {2015})}\BibitemShut {NoStop}%
\bibitem [{\citenamefont {Singer}\ and\ \citenamefont
  {Price}(2016)}]{Singer_2016}%
  \BibitemOpen
  \bibfield  {author} {\bibinfo {author} {\bibfnamefont {L.~P.}\ \bibnamefont
  {Singer}}\ and\ \bibinfo {author} {\bibfnamefont {L.~R.}\ \bibnamefont
  {Price}},\ }\bibfield  {title} {\bibinfo {title} {{Rapid Bayesian position
  reconstruction for gravitational-wave transients}},\ }\href
  {https://doi.org/10.1103/physrevd.93.024013} {\bibfield  {journal} {\bibinfo
  {journal} {Physical Review D}\ }\textbf {\bibinfo {volume} {93}} (\bibinfo
  {year} {2016})}\BibitemShut {NoStop}%
\bibitem [{\citenamefont {LeCun}\ \emph {et~al.}(1989)\citenamefont {LeCun}
  \emph {et~al.}}]{LeCun_1989}%
  \BibitemOpen
  \bibfield  {author} {\bibinfo {author} {\bibfnamefont {Y.}~\bibnamefont
  {LeCun}} \emph {et~al.},\ }\bibfield  {title} {\bibinfo {title}
  {Backpropagation applied to handwritten zip code recognition},\ }\href
  {https://doi.org/10.1162/neco.1989.1.4.541} {\bibfield  {journal} {\bibinfo
  {journal} {Neural Computation}\ }\textbf {\bibinfo {volume} {1}} (\bibinfo
  {year} {1989})}\BibitemShut {NoStop}%
\bibitem [{\citenamefont {LeCun}\ \emph {et~al.}(1998)\citenamefont {LeCun}
  \emph {et~al.}}]{LeCun_1998}%
  \BibitemOpen
  \bibfield  {author} {\bibinfo {author} {\bibfnamefont {Y.}~\bibnamefont
  {LeCun}} \emph {et~al.},\ }\bibfield  {title} {\bibinfo {title}
  {Gradient-based learning applied to document recognition},\ }\href
  {https://doi.org/10.1109/5.726791} {\bibfield  {journal} {\bibinfo  {journal}
  {Proceedings of the {IEEE}}\ }\textbf {\bibinfo {volume} {86}} (\bibinfo
  {year} {1998})}\BibitemShut {NoStop}%
\bibitem [{\citenamefont {Krizhevsky}\ \emph {et~al.}(2012)\citenamefont
  {Krizhevsky}, \citenamefont {Sutskever},\ and\ \citenamefont
  {Hinton}}]{Krizhevsky_2012}%
  \BibitemOpen
  \bibfield  {author} {\bibinfo {author} {\bibfnamefont {A.}~\bibnamefont
  {Krizhevsky}}, \bibinfo {author} {\bibfnamefont {I.}~\bibnamefont
  {Sutskever}},\ and\ \bibinfo {author} {\bibfnamefont {G.~E.}\ \bibnamefont
  {Hinton}},\ }\bibfield  {title} {\bibinfo {title} {Imagenet classification
  with deep convolutional neural networks},\ }in\ \href
  {https://papers.nips.cc/paper/4824-imagenet-classification-with-deep-convolutional-neural-networks.pdf}
  {\emph {\bibinfo {booktitle} {Neural Information Processing Systems
  (NeurIPS)}}}\ (\bibinfo {year} {2012})\BibitemShut {NoStop}%
\bibitem [{\citenamefont {Kim}(2014)}]{Kim_2014}%
  \BibitemOpen
  \bibfield  {author} {\bibinfo {author} {\bibfnamefont {Y.}~\bibnamefont
  {Kim}},\ }\href@noop {} {\bibinfo {title} {Convolutional neural networks for
  sentence classification}} (\bibinfo {year} {2014}),\ \Eprint
  {https://arxiv.org/abs/1408.5882} {arXiv:1408.5882} \BibitemShut {NoStop}%
\bibitem [{\citenamefont {{van den Oord}}\ \emph {et~al.}(2016)\citenamefont
  {{van den Oord}} \emph {et~al.}}]{VanDenOord_2016}%
  \BibitemOpen
  \bibfield  {author} {\bibinfo {author} {\bibfnamefont {A.}~\bibnamefont {{van
  den Oord}}} \emph {et~al.},\ }\href@noop {} {\bibinfo {title} {{WaveNet: A
  Generative Model for Raw Audio}}} (\bibinfo {year} {2016}),\ \Eprint
  {https://arxiv.org/abs/1609.03499} {arXiv:1609.03499} \BibitemShut {NoStop}%
\bibitem [{\citenamefont {Zhu}\ \emph {et~al.}(2014)\citenamefont {Zhu} \emph
  {et~al.}}]{Zhu_2014}%
  \BibitemOpen
  \bibfield  {author} {\bibinfo {author} {\bibfnamefont {W.}~\bibnamefont
  {Zhu}} \emph {et~al.},\ }\bibfield  {title} {\bibinfo {title} {Searching for
  pulsars using image pattern recognition},\ }\href
  {https://doi.org/10.1088/0004-637x/781/2/117} {\bibfield  {journal} {\bibinfo
   {journal} {The Astrophysical Journal}\ }\textbf {\bibinfo {volume} {781}},\
  \bibinfo {pages} {117} (\bibinfo {year} {2014})}\BibitemShut {NoStop}%
\bibitem [{\citenamefont {Carleo}\ \emph {et~al.}(2019)\citenamefont {Carleo}
  \emph {et~al.}}]{Carleo_2019}%
  \BibitemOpen
  \bibfield  {author} {\bibinfo {author} {\bibfnamefont {G.}~\bibnamefont
  {Carleo}} \emph {et~al.},\ }\href@noop {} {\bibinfo {title} {Machine learning
  and the physical sciences}} (\bibinfo {year} {2019}),\ \Eprint
  {https://arxiv.org/abs/1903.10563} {arXiv:1903.10563} \BibitemShut {NoStop}%
\bibitem [{\citenamefont {George}\ and\ \citenamefont
  {Huerta}(2016)}]{George_2016}%
  \BibitemOpen
  \bibfield  {author} {\bibinfo {author} {\bibfnamefont {D.}~\bibnamefont
  {George}}\ and\ \bibinfo {author} {\bibfnamefont {E.~A.}\ \bibnamefont
  {Huerta}},\ }\href@noop {} {\bibinfo {title} {{Deep Neural Networks to Enable
  Real-Time Multimessenger Astrophysics}}} (\bibinfo {year} {2016}),\ \Eprint
  {https://arxiv.org/abs/1701.00008v1} {arXiv:1701.00008v1} \BibitemShut
  {NoStop}%
\bibitem [{\citenamefont {Gabbard}\ \emph {et~al.}(2018)\citenamefont {Gabbard}
  \emph {et~al.}}]{Gabbard_2018}%
  \BibitemOpen
  \bibfield  {author} {\bibinfo {author} {\bibfnamefont {H.}~\bibnamefont
  {Gabbard}} \emph {et~al.},\ }\bibfield  {title} {\bibinfo {title} {Matching
  matched filtering with deep networks for gravitational-wave astronomy},\
  }\href {https://doi.org/10.1103/physrevlett.120.141103} {\bibfield  {journal}
  {\bibinfo  {journal} {Physical Review Letters}\ }\textbf {\bibinfo {volume}
  {120}} (\bibinfo {year} {2018})}\BibitemShut {NoStop}%
\bibitem [{\citenamefont {Nitz}\ \emph {et~al.}(2019)\citenamefont {Nitz} \emph
  {et~al.}}]{PyCBC}%
  \BibitemOpen
  \bibfield  {author} {\bibinfo {author} {\bibfnamefont {A.~H.}\ \bibnamefont
  {Nitz}} \emph {et~al.},\ }\href {https://doi.org/10.5281/zenodo.2581446}
  {\bibinfo {title} {{PyCBC Release v1.13.5}}} (\bibinfo {year}
  {2019})\BibitemShut {NoStop}%
\bibitem [{\citenamefont {Schutz}(1999)}]{Schutz_1999}%
  \BibitemOpen
  \bibfield  {author} {\bibinfo {author} {\bibfnamefont {B.~F.}\ \bibnamefont
  {Schutz}},\ }\bibfield  {title} {\bibinfo {title} {{Gravitational wave
  astronomy}},\ }\href {https://doi.org/10.1088/0264-9381/16/12A/307}
  {\bibfield  {journal} {\bibinfo  {journal} {Classical and Quantum Gravity}\
  }\textbf {\bibinfo {volume} {16}},\ \bibinfo {pages} {A131} (\bibinfo {year}
  {1999})}\BibitemShut {NoStop}%
\bibitem [{\citenamefont {Caudill}(2018)}]{Caudill_2018}%
  \BibitemOpen
  \bibfield  {author} {\bibinfo {author} {\bibfnamefont {S.}~\bibnamefont
  {Caudill}},\ }\bibfield  {title} {\bibinfo {title} {Techniques for
  gravitational-wave detection of compact binary coalescence},\ }in\ \href
  {https://doi.org/10.23919/eusipco.2018.8553549} {\emph {\bibinfo {booktitle}
  {26th European Signal Processing Conference ({EUSIPCO})}}}\ (\bibinfo {year}
  {2018})\BibitemShut {NoStop}%
\bibitem [{\citenamefont {Abbott}\ \emph
  {et~al.}(2016{\natexlab{c}})\citenamefont {Abbott} \emph
  {et~al.}}]{Abbott_2016-06-06}%
  \BibitemOpen
  \bibfield  {author} {\bibinfo {author} {\bibfnamefont {B.~P.}\ \bibnamefont
  {Abbott}} \emph {et~al.},\ }\bibfield  {title} {\bibinfo {title}
  {{Characterization of transient noise in Advanced LIGO relevant to
  gravitational wave signal GW150914}},\ }\href
  {https://doi.org/10.1088/0264-9381/33/13/134001} {\bibfield  {journal}
  {\bibinfo  {journal} {Classical and Quantum Gravity}\ }\textbf {\bibinfo
  {volume} {33}} (\bibinfo {year} {2016}{\natexlab{c}})}\BibitemShut {NoStop}%
\bibitem [{\citenamefont {Cabero}\ \emph {et~al.}(2019)\citenamefont {Cabero}
  \emph {et~al.}}]{Cabero_2019}%
  \BibitemOpen
  \bibfield  {author} {\bibinfo {author} {\bibfnamefont {M.}~\bibnamefont
  {Cabero}} \emph {et~al.},\ }\href@noop {} {\bibinfo {title} {{Blip glitches
  in Advanced LIGO data}}} (\bibinfo {year} {2019}),\ \Eprint
  {https://arxiv.org/abs/1901.05093} {arXiv:1901.05093} \BibitemShut {NoStop}%
\bibitem [{\citenamefont {{LIGO Scientific
  Collaboration}}(2018{\natexlab{a}})}]{LALSuite}%
  \BibitemOpen
  \bibfield  {author} {\bibinfo {author} {\bibnamefont {{LIGO Scientific
  Collaboration}}},\ }\href {https://doi.org/10.7935/GT1W-FZ16} {\bibinfo
  {title} {{LIGO Algorithm Library -- LALSuite}}},\ \bibinfo {howpublished}
  {Free Software (GPL)} (\bibinfo {year} {2018}{\natexlab{a}})\BibitemShut
  {NoStop}%
\bibitem [{\citenamefont {Capano}\ \emph {et~al.}(2016)\citenamefont {Capano}
  \emph {et~al.}}]{Capano_2016}%
  \BibitemOpen
  \bibfield  {author} {\bibinfo {author} {\bibfnamefont {C.}~\bibnamefont
  {Capano}} \emph {et~al.},\ }\bibfield  {title} {\bibinfo {title}
  {Implementing a search for gravitational waves from binary black holes with
  nonprecessing spin},\ }\href {https://doi.org/10.1103/physrevd.93.124007}
  {\bibfield  {journal} {\bibinfo  {journal} {Physical Review D}\ }\textbf
  {\bibinfo {volume} {93}} (\bibinfo {year} {2016})}\BibitemShut {NoStop}%
\bibitem [{\citenamefont {Allen}(2005)}]{Allen_2005}%
  \BibitemOpen
  \bibfield  {author} {\bibinfo {author} {\bibfnamefont {B.}~\bibnamefont
  {Allen}},\ }\bibfield  {title} {\bibinfo {title} {{$\chi^2$~time-frequency
  discriminator for gravitational wave detection}},\ }\href
  {https://doi.org/10.1103/PhysRevD.71.062001} {\bibfield  {journal} {\bibinfo
  {journal} {Physical Review D}\ }\textbf {\bibinfo {volume} {71}} (\bibinfo
  {year} {2005})}\BibitemShut {NoStop}%
\bibitem [{\citenamefont {Nitz}\ \emph {et~al.}(2017)\citenamefont {Nitz} \emph
  {et~al.}}]{Nitz_2017-11-07}%
  \BibitemOpen
  \bibfield  {author} {\bibinfo {author} {\bibfnamefont {A.~H.}\ \bibnamefont
  {Nitz}} \emph {et~al.},\ }\bibfield  {title} {\bibinfo {title} {{Detecting
  Binary Compact-object Mergers with Gravitational Waves: Understanding and
  Improving the Sensitivity of the {PyCBC} Search}},\ }\href
  {https://doi.org/10.3847/1538-4357/aa8f50} {\bibfield  {journal} {\bibinfo
  {journal} {The Astrophysical Journal}\ }\textbf {\bibinfo {volume} {849}},\
  \bibinfo {pages} {118} (\bibinfo {year} {2017})}\BibitemShut {NoStop}%
\bibitem [{\citenamefont {Nitz}(2018)}]{Nitz_2018-01-12}%
  \BibitemOpen
  \bibfield  {author} {\bibinfo {author} {\bibfnamefont {A.~H.}\ \bibnamefont
  {Nitz}},\ }\bibfield  {title} {\bibinfo {title} {Distinguishing short
  duration noise transients in {LIGO} data to improve the {PyCBC} search for
  gravitational waves from high mass binary black hole mergers},\ }\href
  {https://doi.org/10.1088/1361-6382/aaa13d} {\bibfield  {journal} {\bibinfo
  {journal} {Classical and Quantum Gravity}\ }\textbf {\bibinfo {volume} {35}}
  (\bibinfo {year} {2018})}\BibitemShut {NoStop}%
\bibitem [{\citenamefont {LeCun}\ and\ \citenamefont
  {Bengio}(1995)}]{LeCun_1995}%
  \BibitemOpen
  \bibfield  {author} {\bibinfo {author} {\bibfnamefont {Y.}~\bibnamefont
  {LeCun}}\ and\ \bibinfo {author} {\bibfnamefont {Y.}~\bibnamefont {Bengio}},\
  }\bibfield  {title} {\bibinfo {title} {Convolutional networks for images,
  speech, and time-series},\ }in\ \href@noop {} {\emph {\bibinfo {booktitle}
  {The Handbook of Brain Theory and Neural Networks}}}\ (\bibinfo  {publisher}
  {The MIT Press},\ \bibinfo {year} {1995})\BibitemShut {NoStop}%
\bibitem [{\citenamefont {George}\ and\ \citenamefont
  {Huerta}(2018{\natexlab{a}})}]{George_2018a}%
  \BibitemOpen
  \bibfield  {author} {\bibinfo {author} {\bibfnamefont {D.}~\bibnamefont
  {George}}\ and\ \bibinfo {author} {\bibfnamefont {E.~A.}\ \bibnamefont
  {Huerta}},\ }\bibfield  {title} {\bibinfo {title} {Deep neural networks to
  enable real-time multimessenger astrophysics},\ }\href
  {https://doi.org/10.1103/physrevd.97.044039} {\bibfield  {journal} {\bibinfo
  {journal} {Physical Review D}\ }\textbf {\bibinfo {volume} {97}} (\bibinfo
  {year} {2018}{\natexlab{a}})}\BibitemShut {NoStop}%
\bibitem [{\citenamefont {George}\ and\ \citenamefont
  {Huerta}(2018{\natexlab{b}})}]{George_2018b}%
  \BibitemOpen
  \bibfield  {author} {\bibinfo {author} {\bibfnamefont {D.}~\bibnamefont
  {George}}\ and\ \bibinfo {author} {\bibfnamefont {E.~A.}\ \bibnamefont
  {Huerta}},\ }\bibfield  {title} {\bibinfo {title} {{Deep Learning for
  real-time gravitational wave detection and parameter estimation: Results with
  Advanced LIGO data}},\ }\href
  {https://doi.org/10.1016/j.physletb.2017.12.053} {\bibfield  {journal}
  {\bibinfo  {journal} {Physics Letters B}\ }\textbf {\bibinfo {volume} {778}}
  (\bibinfo {year} {2018}{\natexlab{b}})}\BibitemShut {NoStop}%
\bibitem [{\citenamefont {Li}\ \emph {et~al.}(2017)\citenamefont {Li},
  \citenamefont {Yu},\ and\ \citenamefont {Fan}}]{Li_2017}%
  \BibitemOpen
  \bibfield  {author} {\bibinfo {author} {\bibfnamefont {X.}~\bibnamefont
  {Li}}, \bibinfo {author} {\bibfnamefont {W.}~\bibnamefont {Yu}},\ and\
  \bibinfo {author} {\bibfnamefont {X.}~\bibnamefont {Fan}},\ }\href@noop {}
  {\bibinfo {title} {A method of detecting gravitational wave based on
  time-frequency analysis and convolutional neural networks}} (\bibinfo {year}
  {2017}),\ \Eprint {https://arxiv.org/abs/1712.00356} {arXiv:1712.00356}
  \BibitemShut {NoStop}%
\bibitem [{\citenamefont {Zevin}\ \emph {et~al.}(2017)\citenamefont {Zevin}
  \emph {et~al.}}]{Zevin_2017}%
  \BibitemOpen
  \bibfield  {author} {\bibinfo {author} {\bibfnamefont {M.}~\bibnamefont
  {Zevin}} \emph {et~al.},\ }\bibfield  {title} {\bibinfo {title} {{Gravity
  Spy: integrating advanced {LIGO} detector characterization, machine learning,
  and citizen science}},\ }\href {https://doi.org/10.1088/1361-6382/aa5cea}
  {\bibfield  {journal} {\bibinfo  {journal} {Classical and Quantum Gravity}\
  }\textbf {\bibinfo {volume} {34}} (\bibinfo {year} {2017})}\BibitemShut
  {NoStop}%
\bibitem [{\citenamefont {Bahaadini}\ \emph {et~al.}(2017)\citenamefont
  {Bahaadini} \emph {et~al.}}]{Bahaadini_2017}%
  \BibitemOpen
  \bibfield  {author} {\bibinfo {author} {\bibfnamefont {S.}~\bibnamefont
  {Bahaadini}} \emph {et~al.},\ }\bibfield  {title} {\bibinfo {title} {Deep
  multi-view models for glitch classification},\ }in\ \href
  {https://doi.org/10.1109/icassp.2017.7952693} {\emph {\bibinfo {booktitle}
  {{International Conference on Acoustics, Speech and Signal Processing
  ({ICASSP})}}}}\ (\bibinfo {year} {2017})\BibitemShut {NoStop}%
\bibitem [{\citenamefont {Razzano}\ and\ \citenamefont
  {Cuoco}(2018)}]{Razzano_2018}%
  \BibitemOpen
  \bibfield  {author} {\bibinfo {author} {\bibfnamefont {M.}~\bibnamefont
  {Razzano}}\ and\ \bibinfo {author} {\bibfnamefont {E.}~\bibnamefont
  {Cuoco}},\ }\bibfield  {title} {\bibinfo {title} {Image-based deep learning
  for classification of noise transients in gravitational wave detectors},\
  }\href {https://doi.org/10.1088/1361-6382/aab793} {\bibfield  {journal}
  {\bibinfo  {journal} {Classical and Quantum Gravity}\ }\textbf {\bibinfo
  {volume} {35}} (\bibinfo {year} {2018})}\BibitemShut {NoStop}%
\bibitem [{\citenamefont {Bahaadini}\ \emph {et~al.}(2018)\citenamefont
  {Bahaadini} \emph {et~al.}}]{Bahaadini_2018}%
  \BibitemOpen
  \bibfield  {author} {\bibinfo {author} {\bibfnamefont {S.}~\bibnamefont
  {Bahaadini}} \emph {et~al.},\ }\bibfield  {title} {\bibinfo {title} {{Machine
  learning for Gravity Spy: Glitch classification and dataset}},\ }\href
  {https://doi.org/10.1016/j.ins.2018.02.068} {\bibfield  {journal} {\bibinfo
  {journal} {Information Sciences}\ }\textbf {\bibinfo {volume} {444}}
  (\bibinfo {year} {2018})}\BibitemShut {NoStop}%
\bibitem [{\citenamefont {Coughlin}\ \emph {et~al.}(2019)\citenamefont
  {Coughlin} \emph {et~al.}}]{Coughlin_2019}%
  \BibitemOpen
  \bibfield  {author} {\bibinfo {author} {\bibfnamefont {S.~B.}\ \bibnamefont
  {Coughlin}} \emph {et~al.},\ }\href@noop {} {\bibinfo {title} {Classifying
  the unknown: discovering novel gravitational-wave detector glitches using
  similarity learning}} (\bibinfo {year} {2019}),\ \Eprint
  {https://arxiv.org/abs/1903.04058} {arXiv:1903.04058} \BibitemShut {NoStop}%
\bibitem [{\citenamefont {Shen}\ \emph {et~al.}(2019)\citenamefont {Shen},
  \citenamefont {Huerta},\ and\ \citenamefont {Zhao}}]{Shen_2019}%
  \BibitemOpen
  \bibfield  {author} {\bibinfo {author} {\bibfnamefont {H.}~\bibnamefont
  {Shen}}, \bibinfo {author} {\bibfnamefont {E.~A.}\ \bibnamefont {Huerta}},\
  and\ \bibinfo {author} {\bibfnamefont {Z.}~\bibnamefont {Zhao}},\ }\href@noop
  {} {\bibinfo {title} {{Deep Learning at Scale for Gravitational Wave
  Parameter Estimation of Binary Black Hole Mergers}}} (\bibinfo {year}
  {2019}),\ \Eprint {https://arxiv.org/abs/1903.01998} {arXiv:1903.01998}
  \BibitemShut {NoStop}%
\bibitem [{\citenamefont {Dreissigacker}\ \emph {et~al.}(2019)\citenamefont
  {Dreissigacker} \emph {et~al.}}]{Dreissigacker_2019}%
  \BibitemOpen
  \bibfield  {author} {\bibinfo {author} {\bibfnamefont {C.}~\bibnamefont
  {Dreissigacker}} \emph {et~al.},\ }\href@noop {} {\bibinfo {title}
  {{Deep-Learning Continuous Gravitational Waves}}} (\bibinfo {year} {2019}),\
  \Eprint {https://arxiv.org/abs/1904.13291} {arXiv:1904.13291} \BibitemShut
  {NoStop}%
\bibitem [{\citenamefont {Nitz}\ \emph {et~al.}(2018)\citenamefont {Nitz} \emph
  {et~al.}}]{Nitz_2018-07-30}%
  \BibitemOpen
  \bibfield  {author} {\bibinfo {author} {\bibfnamefont {A.~H.}\ \bibnamefont
  {Nitz}} \emph {et~al.},\ }\bibfield  {title} {\bibinfo {title} {{Rapid
  detection of gravitational waves from compact binary mergers with {PyCBC}
  Live}},\ }\href {https://doi.org/10.1103/physrevd.98.024050} {\bibfield
  {journal} {\bibinfo  {journal} {Physical Review D}\ }\textbf {\bibinfo
  {volume} {98}} (\bibinfo {year} {2018})}\BibitemShut {NoStop}%
\bibitem [{\citenamefont {Gebhard}\ and\ \citenamefont
  {Kilbertus}(2019{\natexlab{a}})}]{Gebhard_2019a}%
  \BibitemOpen
  \bibfield  {author} {\bibinfo {author} {\bibfnamefont {T.~D.}\ \bibnamefont
  {Gebhard}}\ and\ \bibinfo {author} {\bibfnamefont {N.}~\bibnamefont
  {Kilbertus}},\ }\href {https://doi.org/10.5281/zenodo.2649358} {\bibinfo
  {title} {ggwd: generate gravi\-tational-wave data}} (\bibinfo {year}
  {2019}{\natexlab{a}}),\ \bibinfo {note} {{DOI:}
  \texttt{10.5281/zenodo.2649358}}\BibitemShut {NoStop}%
\bibitem [{\citenamefont {Abbott}\ \emph
  {et~al.}(2016{\natexlab{d}})\citenamefont {Abbott} \emph
  {et~al.}}]{Abbott_2016-03-31}%
  \BibitemOpen
  \bibfield  {author} {\bibinfo {author} {\bibfnamefont {B.~P.}\ \bibnamefont
  {Abbott}} \emph {et~al.},\ }\bibfield  {title} {\bibinfo {title} {{GW150914:
  The Advanced LIGO Detectors in the Era of First Discoveries}},\ }\href
  {https://doi.org/10.1103/physrevlett.116.131103} {\bibfield  {journal}
  {\bibinfo  {journal} {Physical Review Letters}\ }\textbf {\bibinfo {volume}
  {116}} (\bibinfo {year} {2016}{\natexlab{d}})}\BibitemShut {NoStop}%
\bibitem [{\citenamefont {Martynov}\ \emph {et~al.}(2016)\citenamefont
  {Martynov} \emph {et~al.}}]{Martynov_2016}%
  \BibitemOpen
  \bibfield  {author} {\bibinfo {author} {\bibfnamefont {D.~V.}\ \bibnamefont
  {Martynov}} \emph {et~al.},\ }\bibfield  {title} {\bibinfo {title}
  {Sensitivity of the advanced {LIGO} detectors at the beginning of
  gravitational wave astronomy},\ }\href
  {https://doi.org/10.1103/physrevd.93.112004} {\bibfield  {journal} {\bibinfo
  {journal} {Physical Review D}\ }\textbf {\bibinfo {volume} {93}} (\bibinfo
  {year} {2016})}\BibitemShut {NoStop}%
\bibitem [{\citenamefont {Abbott}\ \emph
  {et~al.}(2017{\natexlab{g}})\citenamefont {Abbott} \emph
  {et~al.}}]{Abbott_2017-03-28}%
  \BibitemOpen
  \bibfield  {author} {\bibinfo {author} {\bibfnamefont {B.~P.}\ \bibnamefont
  {Abbott}} \emph {et~al.},\ }\bibfield  {title} {\bibinfo {title}
  {{Calibration of the Advanced LIGO detectors for the discovery of the binary
  black-hole merger GW150914}},\ }\href
  {https://doi.org/10.1103/physrevd.95.062003} {\bibfield  {journal} {\bibinfo
  {journal} {Physical Review D}\ }\textbf {\bibinfo {volume} {95}} (\bibinfo
  {year} {2017}{\natexlab{g}})}\BibitemShut {NoStop}%
\bibitem [{\citenamefont {Vallisneri}\ \emph {et~al.}(2015)\citenamefont
  {Vallisneri} \emph {et~al.}}]{Vallisneri_2015}%
  \BibitemOpen
  \bibfield  {author} {\bibinfo {author} {\bibfnamefont {M.}~\bibnamefont
  {Vallisneri}} \emph {et~al.},\ }\bibfield  {title} {\bibinfo {title} {{The
  {LIGO} Open Science Center}},\ }\href
  {https://doi.org/10.1088/1742-6596/610/1/012021} {\bibfield  {journal}
  {\bibinfo  {journal} {Journal of Physics: Conference Series}\ }\textbf
  {\bibinfo {volume} {610}} (\bibinfo {year} {2015})}\BibitemShut {NoStop}%
\bibitem [{\citenamefont {{LIGO Scientific
  Collaboration}}(2018{\natexlab{b}})}]{GWOSC_2018}%
  \BibitemOpen
  \bibfield  {author} {\bibinfo {author} {\bibnamefont {{LIGO Scientific
  Collaboration}}},\ }\href {https://doi.org/10.7935/k57p8w9d} {\bibinfo
  {title} {{O1 Data Release}}} (\bibinfo {year}
  {2018}{\natexlab{b}})\BibitemShut {NoStop}%
\bibitem [{\citenamefont {Biwer}\ \emph {et~al.}(2017)\citenamefont {Biwer}
  \emph {et~al.}}]{Biwer_2017}%
  \BibitemOpen
  \bibfield  {author} {\bibinfo {author} {\bibfnamefont {C.}~\bibnamefont
  {Biwer}} \emph {et~al.},\ }\bibfield  {title} {\bibinfo {title} {{Validating
  gravitational-wave detections: The Advanced {LIGO} hardware injection
  system}},\ }\href {https://doi.org/10.1103/physrevd.95.062002} {\bibfield
  {journal} {\bibinfo  {journal} {Physical Review D}\ }\textbf {\bibinfo
  {volume} {95}} (\bibinfo {year} {2017})}\BibitemShut {NoStop}%
\bibitem [{\citenamefont {Bohé}\ \emph {et~al.}(2017)\citenamefont {Bohé}
  \emph {et~al.}}]{Bohe_2017}%
  \BibitemOpen
  \bibfield  {author} {\bibinfo {author} {\bibfnamefont {A.}~\bibnamefont
  {Bohé}} \emph {et~al.},\ }\bibfield  {title} {\bibinfo {title} {Improved
  effective-one-body model of spinning, nonprecessing binary black holes for
  the era of gravitational-wave astrophysics with advanced detectors},\ }\href
  {https://doi.org/10.1103/PhysRevD.95.044028} {\bibfield  {journal} {\bibinfo
  {journal} {Physical Review D}\ }\textbf {\bibinfo {volume} {95}} (\bibinfo
  {year} {2017})}\BibitemShut {NoStop}%
\bibitem [{\citenamefont {Bengio}\ \emph {et~al.}(2009)\citenamefont {Bengio}
  \emph {et~al.}}]{Bengio_2009}%
  \BibitemOpen
  \bibfield  {author} {\bibinfo {author} {\bibfnamefont {Y.}~\bibnamefont
  {Bengio}} \emph {et~al.},\ }\bibfield  {title} {\bibinfo {title} {{Curriculum
  Learning}},\ }in\ \href {https://doi.org/10.1145/1553374.1553380} {\emph
  {\bibinfo {booktitle} {{International Conference on Machine Learning
  (ICML)}}}}\ (\bibinfo {year} {2009})\BibitemShut {NoStop}%
\bibitem [{Note1()}]{Note1}%
  \BibitemOpen
  \bibinfo {note} {For further details, see, for example, section 9.5.3 (b) in
  \cite {Thorne_1987}, section 9.2.3 in \cite {Maggiore_2008}, or section
  6.1.11 in \cite {Creighton_2011}.}\BibitemShut {Stop}%
\bibitem [{\citenamefont {Cutler}\ and\ \citenamefont
  {Flanagan}(1994)}]{Cutler_1994}%
  \BibitemOpen
  \bibfield  {author} {\bibinfo {author} {\bibfnamefont {C.}~\bibnamefont
  {Cutler}}\ and\ \bibinfo {author} {\bibfnamefont {{\'E}.~E.}\ \bibnamefont
  {Flanagan}},\ }\bibfield  {title} {\bibinfo {title} {Gravitational waves from
  merging compact binaries: How accurately can one extract the binary's
  parameters from the inspiral waveform?},\ }\href
  {https://doi.org/10.1103/physrevd.49.2658} {\bibfield  {journal} {\bibinfo
  {journal} {Physical Review D}\ }\textbf {\bibinfo {volume} {49}} (\bibinfo
  {year} {1994})}\BibitemShut {NoStop}%
\bibitem [{\citenamefont {Smith}\ \emph {et~al.}(2013)\citenamefont {Smith},
  \citenamefont {Mandel},\ and\ \citenamefont {Vecchio}}]{Smith_2013}%
  \BibitemOpen
  \bibfield  {author} {\bibinfo {author} {\bibfnamefont {R.~J.~E.}\
  \bibnamefont {Smith}}, \bibinfo {author} {\bibfnamefont {I.}~\bibnamefont
  {Mandel}},\ and\ \bibinfo {author} {\bibfnamefont {A.}~\bibnamefont
  {Vecchio}},\ }\bibfield  {title} {\bibinfo {title} {Studies of waveform
  requirements for intermediate mass-ratio coalescence searches with advanced
  gravitational-wave detectors},\ }\href
  {https://doi.org/10.1103/physrevd.88.044010} {\bibfield  {journal} {\bibinfo
  {journal} {Physical Review D}\ }\textbf {\bibinfo {volume} {88}} (\bibinfo
  {year} {2013})}\BibitemShut {NoStop}%
\bibitem [{Note2()}]{Note2}%
  \BibitemOpen
  \bibinfo {note} {The sigmoid activation function $\sigma : \protect \mathbb
  {R} \to (0,1)$ is defined as $\sigma (x) := 1 / (1 + e^{-x})$.}\BibitemShut
  {Stop}%
\bibitem [{\citenamefont {Paszke}\ \emph {et~al.}(2017)\citenamefont {Paszke}
  \emph {et~al.}}]{Paszke_2017}%
  \BibitemOpen
  \bibfield  {author} {\bibinfo {author} {\bibfnamefont {A.}~\bibnamefont
  {Paszke}} \emph {et~al.},\ }\bibfield  {title} {\bibinfo {title} {{Automatic
  differentiation in PyTorch}},\ }\href {https://autodiff-workshop.github.io/}
  {\bibfield  {journal} {\bibinfo  {journal} {Accepted at the ``Autodiff
  Workshop'' at NeurIPS}\ } (\bibinfo {year} {2017})}\BibitemShut {NoStop}%
\bibitem [{\citenamefont {Gebhard}\ and\ \citenamefont
  {Kilbertus}(2019{\natexlab{b}})}]{Gebhard_2019b}%
  \BibitemOpen
  \bibfield  {author} {\bibinfo {author} {\bibfnamefont {T.~D.}\ \bibnamefont
  {Gebhard}}\ and\ \bibinfo {author} {\bibfnamefont {N.}~\bibnamefont
  {Kilbertus}},\ }\href {https://doi.org/10.5281/zenodo.2649433} {\bibinfo
  {title} {{Source code for ``Convolutional neural networks: a magic bullet for
  gravitational-wave detection?''}}} (\bibinfo {year} {2019}{\natexlab{b}}),\
  \bibinfo {note} {{DOI:} \texttt{10.5281/zenodo.2649433}}\BibitemShut
  {NoStop}%
\bibitem [{\citenamefont {He}\ \emph {et~al.}(2015)\citenamefont {He} \emph
  {et~al.}}]{He_2015}%
  \BibitemOpen
  \bibfield  {author} {\bibinfo {author} {\bibfnamefont {K.}~\bibnamefont {He}}
  \emph {et~al.},\ }\href@noop {} {\bibinfo {title} {{Delving Deep into
  Rectifiers: Surpassing Human-Level Performance on ImageNet Classification}}}
  (\bibinfo {year} {2015}),\ \Eprint {https://arxiv.org/abs/1502.01852}
  {arXiv:1502.01852} \BibitemShut {NoStop}%
\bibitem [{\citenamefont {Kingma}\ and\ \citenamefont
  {Ba}(2014)}]{Kingma_2014}%
  \BibitemOpen
  \bibfield  {author} {\bibinfo {author} {\bibfnamefont {D.~P.}\ \bibnamefont
  {Kingma}}\ and\ \bibinfo {author} {\bibfnamefont {J.}~\bibnamefont {Ba}},\
  }\href@noop {} {\bibinfo {title} {{Adam: A method for stochastic
  optimization}}} (\bibinfo {year} {2014}),\ \Eprint
  {https://arxiv.org/abs/1412.6980} {arXiv:1412.6980} \BibitemShut {NoStop}%
\bibitem [{\citenamefont {Reddi}\ \emph {et~al.}(2018)\citenamefont {Reddi},
  \citenamefont {Kale},\ and\ \citenamefont {Kumar}}]{Reddi_2018}%
  \BibitemOpen
  \bibfield  {author} {\bibinfo {author} {\bibfnamefont {S.~J.}\ \bibnamefont
  {Reddi}}, \bibinfo {author} {\bibfnamefont {S.}~\bibnamefont {Kale}},\ and\
  \bibinfo {author} {\bibfnamefont {S.}~\bibnamefont {Kumar}},\ }\bibfield
  {title} {\bibinfo {title} {{On the convergence of Adam and beyond}},\ }in\
  \href {https://openreview.net/pdf?id=ryQu7f-RZ} {\emph {\bibinfo {booktitle}
  {International Conference on Learning Representations (ICLR)}}}\ (\bibinfo
  {year} {2018})\BibitemShut {NoStop}%
\bibitem [{\citenamefont {Zeiler}\ and\ \citenamefont
  {Fergus}(2014)}]{Zeiler_2014}%
  \BibitemOpen
  \bibfield  {author} {\bibinfo {author} {\bibfnamefont {M.~D.}\ \bibnamefont
  {Zeiler}}\ and\ \bibinfo {author} {\bibfnamefont {R.}~\bibnamefont
  {Fergus}},\ }\bibfield  {title} {\bibinfo {title} {Visualizing and
  understanding convolutional networks},\ }in\ \href
  {https://doi.org/10.1007/978-3-319-10590-1_53} {\emph {\bibinfo {booktitle}
  {European Conference on Computer Vision (ECCV)}}}\ (\bibinfo {year} {2014})\
  pp.\ \bibinfo {pages} {818--833}\BibitemShut {NoStop}%
\bibitem [{\citenamefont {Olah}\ \emph {et~al.}(2017)\citenamefont {Olah},
  \citenamefont {Mordvintsev},\ and\ \citenamefont {Schubert}}]{Olah_2017}%
  \BibitemOpen
  \bibfield  {author} {\bibinfo {author} {\bibfnamefont {C.}~\bibnamefont
  {Olah}}, \bibinfo {author} {\bibfnamefont {A.}~\bibnamefont {Mordvintsev}},\
  and\ \bibinfo {author} {\bibfnamefont {L.}~\bibnamefont {Schubert}},\
  }\bibfield  {title} {\bibinfo {title} {Feature visualization},\ }\href
  {https://doi.org/10.23915/distill.00007} {\bibfield  {journal} {\bibinfo
  {journal} {Distill}\ }\textbf {\bibinfo {volume} {2}} (\bibinfo {year}
  {2017})}\BibitemShut {NoStop}%
\bibitem [{\citenamefont {Szegedy}\ \emph {et~al.}(2013)\citenamefont {Szegedy}
  \emph {et~al.}}]{Szegedy_2013}%
  \BibitemOpen
  \bibfield  {author} {\bibinfo {author} {\bibfnamefont {C.}~\bibnamefont
  {Szegedy}} \emph {et~al.},\ }\href@noop {} {\bibinfo {title} {Intriguing
  properties of neural networks}} (\bibinfo {year} {2013}),\ \Eprint
  {https://arxiv.org/abs/1312.6199} {arXiv:1312.6199} \BibitemShut {NoStop}%
\bibitem [{\citenamefont {Thorne}(1987)}]{Thorne_1987}%
  \BibitemOpen
  \bibfield  {author} {\bibinfo {author} {\bibfnamefont {K.~S.}\ \bibnamefont
  {Thorne}},\ }\bibfield  {title} {\bibinfo {title} {{Gravitational
  radiation}},\ }in\ \href@noop {} {\emph {\bibinfo {booktitle} {{Three hundred
  years of gravitation}}}},\ \bibinfo {editor} {edited by\ \bibinfo {editor}
  {\bibfnamefont {S.~W.}\ \bibnamefont {Hawking}}\ and\ \bibinfo {editor}
  {\bibfnamefont {W.}~\bibnamefont {Israel}}}\ (\bibinfo  {publisher}
  {Cambridge University Press},\ \bibinfo {address} {Cambridge},\ \bibinfo
  {year} {1987})\ pp.\ \bibinfo {pages} {330--458}\BibitemShut {NoStop}%
\bibitem [{\citenamefont {Maggiore}(2008)}]{Maggiore_2008}%
  \BibitemOpen
  \bibfield  {author} {\bibinfo {author} {\bibfnamefont {M.}~\bibnamefont
  {Maggiore}},\ }\href@noop {} {\emph {\bibinfo {title} {{Gravitational Waves:
  Volume 1: Theory and Experiments}}}}\ (\bibinfo  {publisher} {Oxford
  University Press},\ \bibinfo {address} {New York},\ \bibinfo {year}
  {2008})\BibitemShut {NoStop}%
\bibitem [{\citenamefont {Creighton}\ and\ \citenamefont
  {Anderson}(2011)}]{Creighton_2011}%
  \BibitemOpen
  \bibfield  {author} {\bibinfo {author} {\bibfnamefont {J.~D.~E.}\
  \bibnamefont {Creighton}}\ and\ \bibinfo {author} {\bibfnamefont {W.~G.}\
  \bibnamefont {Anderson}},\ }\href@noop {} {\emph {\bibinfo {title}
  {Gravitational-Wave Physics and Astronomy: An Introduction to Theory,
  Experiment and Data Analysis}}}\ (\bibinfo  {publisher} {Wiley-VCH},\
  \bibinfo {address} {Weinheim},\ \bibinfo {year} {2011})\BibitemShut {NoStop}%
\end{thebibliography}%

\end{document}